# Does shot noise always provide the quasiparticle charge?

Sourav Biswas[1], Rajarshi Bhattacharyya[1], Hemanta Kumar Kundu[1], Ankur Das[1], Moty Heiblum[1*], Vladimir Umansky[1], Moshe Goldstein[2] and Yuval Gefen[1]

[1]*Braun Center for Submicron Research, Department of Condensed Matter Physics, Weizmann Institute of Science, Rehovot 7610001, Israel*

[2]*Raymond and Beverly Sackler School of Physics and Astronomy, Tel-Aviv University, Tel Aviv 6997801, Israel*

[*]e-mail: moty.heiblum@weizmann.ac.il

**The fractional charge of quasiparticles is a fundamental feature of quantum Hall effect (QHE) States. The charge has long been measured via shot-noise at moderate temperatures (*T*>30mK), with the Fano factor $F=e^*/e$ revealing the charge $e^*$ of the quasiparticles[1-3]. However, at sufficiently low temperatures (*T*≈10mK), we consistently find *F* being equal to the bulk filling factor, $\nu_b$[4]. Surprisingly, noise with $F = \nu_b$ is also observed on intermediate conductance plateaus in the transmission of the quantum point contact (QPC), where shot noise is not expected[5,6]. We attribute the unexpected Fano factor to upstream neutral modes, which proliferate at the lowest spinless Landau level[6,7]. The universality of the Fano factor is also confirmed when the edge modes do not conform to the bulk. For this, the ubiquitous edge modes at the periphery of the sample are replaced by artificially constructed 'interface modes', propagating at the interface between two adjoined QHE states: the tested state and a different state. We present a new theoretical paradigm based on an interplay between charge and neutral modes, explaining the origin of the universal Fano factor.**

## I. Introduction

The quantum Hall effect (QHE) is the oldest known topological quantum system[8,9]. The most natural description of its gapped bulk is understood via the so-called Landau levels (LLs), which host localized quasiparticles, with different topological phases described by the filling factor (ν). While the bulk is insulating, the low energy dynamics is governed by gapless chiral edge modes[10-13], which carry current and energy. 'Bulk-Edge' correspondence dictates that both the Hall conductance $\sigma_{xy}$ and the thermal Hall conductance $K_{xy}$ are determined by the topological order of the bulk[14]. The quasiparticle charge, $e^*$, being the most fundamental quantity of a fractional state, is crucial in determining the quasiparticle statistics[15].

According to the orthodox paradigm, weak partitioning of edge modes, done in a quantum point contact (QPC) constriction, leads to shot noise with a Fano factor $F=e^*/e$ ($F$=1 for integers)[16-18]. Moreover, shot noise (at zero temperature) should be zero on an intermediate conductance plateau of the QPC, as modes are either fully transmitted or fully reflected, i.e., no intra-mode partitioning. Indeed, early on, experimental results performed at moderate temperatures, $T$>30mK adhered to this expectation[2,3,19,20].

However, in more recent shot noise measurements, performed at $T$≈10mK, we found[4-6]: ***i***. Non-zero noise on intermediate conductance plateaus with the partitioning QPC; ***ii***. The Fano factor is equal to the filling factor in the bulk (away from the QPC), on and off the intermediate conductance plateaus. These observations necessitate further investigation and a new understanding beyond the orthodox paradigm.

To demonstrate the universality of the Fano factor, here we employed a novel method in which two different states are interfaced, giving birth to 1D 'interface modes' at the boundary between the two states. These modes do not obey 'bulk – edge' correspondence, having a different 'effective filling' (conductance) than the filling within the QPC constriction.

It was already demonstrated that *upstream* (US) *neutral* edge modes proliferate in many QHE states in the lowest LL[6,7,21]. These modes can be topological (determined by the bulk) or emergent due to spontaneous edge-reconstruction[7,22]. Here, we show that neutral modes play a crucial role in generating (partitioned) shot noise. The new paradigm of shot noise generation consists of a two-step process: inter-mode charge equilibration accompanied by the generation of neutral quasi-particles and the subsequent annihilation of these "neutralons", leading to stochastic generation of quasi-particle/quasi-hole pairs, resulting in shot noise. We show that the latter is characterized by a Fano factor that is equal to the bulk filling factor. Notably, the Fano factor does not depend on the edge mode structure (and its conductance) and not on the local filling within the QPC constriction (if not too small).

## II. Past results and new measurement platform

**Past results:** Early shot noise measurements, performed at $v$=1/3 and $v$=2/5 'particle states' at moderate temperatures ($T$>25mK), led to $F$≅1/3 and $F$≅1/5, respectively[2,3]. However, later measurements with $\nu$=2/5, $\nu$=3/7, performed at lower temperatures ($T$<15mK), approached $F\cong\nu_b$[4]. Increasing the temperature led to $F\cong$1/5 and $F\cong$1/7, respectively. Similarly, low-temperature measurements in 'particle-hole conjugated' states, $\nu$=2/3, $\nu$=3/5, and $\nu$=4/7, measured 'on' and 'off' the QPC intermediate plateaus, resulted in $F\cong\nu_b$[5,6]. At $\nu$=2/3, with an increased temperature, $F\cong$1/3[5] with $e^*=e/3$ quasi-particle charge. Observation of noise on an intermediate conductance plateau necessitates overhauling the existing paradigm of shot noise.

Consider first the example of $\nu$=1/3 bulk filling. The conventional picture of the edge structure comprises one downstream (DS) charged mode. However, edge-reconstruction[23], taking place when the edge confining potential is not steep, may lead to pairs of additional $\nu$=1/m counter-propagating modes (m an odd integer). This is similar to the reconstruction of the $\nu$=2/3 edge[24,25] and the $\nu$=1 edge[6,26] with m=3[5,27]. Accounting for an interplay of inter-mode Coulomb interactions and disorder-induced tunneling, a new 'fixed point' that features neutral mode(s) settles in[22,28]: two DS charge modes and one US neutral mode, keeping $K_{xy}$ unchanged. In a similar way, with edge reconstruction, $\nu$=2/5 is constituted of three DS charged modes and one US neutral mode. The unreconstructed edge will support two conventional DS charge modes, an inner $\nu$=1/15 and an outer $\nu$=1/3. Our paradigm for the 'reconstructed shot noise' relies on an excited US neutral modes, due to charge equilibration between the 'hot' charge modes that emanate from the biased source and the 'cold' charge modes emanating from the grounded contact[26,27]. Moving US, the neutral modes decay, generating randomized charged quasiparticle/quasihole pairs on different charge modes. Each of the pairs split, with the quasiparticles (quasiholes) arriving at drain D1 (where the noise is measured, see Fig 2a) and the quasiholes (quasiparticles) at drain D2. This adds a stochastic component to the current measured in D1.

The mechanism described here (for details, see below), is going beyond the partitioned beam paradigm for shot noise. It only assumes complete decay of the excited neutral modes in the course of equilibration. The number of those (equal to the number of stochastically generated quasiparticle/quasihole pairs) is defined by the filling of the bulk away from the QPC.

**The new platform:** To reaffirm further the universality of the Fano factor, we exploited a fabrication method allowing to interface the tested state with bulk filling $\nu_b$ with an adjacent (gate-controlled) state $\nu_g$. The resultant interface mode, with an effective filling '$\nu_b$-$\nu_g$' was partitioned by a QPC with its own shot noise, with 'bulk-edge' correspondence not valid. This shot noise was compared with the ubiquitous shot noise of partitioned edge modes.

Our playground was a standard MBE grown GaAs-AlGaAs heterostructures harboring high mobility 2DEG, located 86 nm below the surface, with an electron density $1.7 \times 10^{11}\mathrm{cm}^{-2}$. Electrical measurements were carried out at electron temperature $\sim 12 - 14$ mK.

The deployed platform is shown in Fig. 1a. A gate-defined Hall bar, with filling $\nu_\mathrm{g}$ (green) is embedded in the bulk of a mesa having a filling $\nu_\mathrm{b}$ (light blue). If $\nu_\mathrm{g} < \nu_\mathrm{b}$, an equilibrium *interface mode* circulates the Hall bar with an 'effective filling' $\nu_\mathrm{int} = \nu_\mathrm{b} - \nu_\mathrm{g}$ and conductance $G = \nu_\mathrm{int}\frac{e^2}{h}$, measured between ohmic contacts (marked yellow) placed at the formed interface. In the example of Fig. 1a, with $\nu_\mathrm{int} = 2 - 1 = 1$, only the inner integer mode participates in the transport.

Figure 1b shows the Hall resistance $R_\mathrm{xy}$ of the gated region as a function of the gate voltage $V_\mathrm{g}$, when $\nu_\mathrm{b} = 2$. As $V_\mathrm{g} \to 0\mathrm{V}$, $R_\mathrm{xy} \to \infty$ (the interface vanishes). For $V_\mathrm{g} < -0.35\mathrm{V}$, the gated region is fully depleted and $R_\mathrm{xy} = \frac{h}{2e^2}$. Five conductance plateaus in units of $\frac{e^2}{h}$ are observed: $\frac{1}{3}$, $\frac{2}{3}, 1, \frac{4}{3}, \frac{5}{3}, 2$, assuring charge equilibration[29] at the interface.

**Noise measurements:** Shot noise was measured by partitioning the interface modes (Fig. 2a). The QPC constriction was formed by the lower and upper gated regions. The spectral density of the charge fluctuations were filtered by a LC resonant-circuit with a center frequency ~937kHz and bandwidth of 30kHz. A HEMT-based preamplifier, cooled to ~4K (voltage noise ~300pV/√Hz and current noise ~10fA/√Hz), was cascaded by a room temperature amplifier (with voltage noise 0.5nV/√Hz), feeding a spectrum analyzer.

The general expression for the low-frequency spectral density of the shot noise is: $S_i(0) = 2FeI_\mathrm{dc}t(1-t)\{\coth\left(\frac{FeV}{2k_\mathrm{B}T}\right) - \frac{2k_\mathrm{B}T}{FeV}\}$, with $t$ the transmission of the QPC and $I_\mathrm{dc} = V_\mathrm{S}G$ the impinging current[5,30-33]. The gain is calibrated by measuring the shot noise at weak backscattering of the outer edge mode in the '2-0' configuration, with an expected $F$=1. The detailed procedure of the analysis is described in the Extended Data (ED) Section-I.

## III. New measurement results

We generalize the ubiquitous noise measurements by presenting key results of partitioning interface modes. We control the lower (upper) interface mode with an effective filling, $\nu_{\mathrm{int}}^{l}$ ($\nu_{\mathrm{int}}^{u}$) (see Fig. 2a), with bulk filling $\nu_{\mathrm{b}}$. We test: ***i.*** Different (incoming) modes' filling at a fixed bulk filling; ***ii.*** A fixed incoming mode filling at different bulk filling.

***Fixed*** $\boldsymbol{\nu_{\mathrm{b}}}$**=1:** Conductance showing fractional modes at the lower interface is plotted as a function of the lower gate voltage, $V_{\mathrm{g}}^{l}$ (Fig. 2b). An example of tuning the QPC transmission with $V_{\mathrm{g}}^{u}$ when the lower gated region is depleted, and $\nu_{\mathrm{int}}^{l}=1$ is shown in Fig. 2c. Here is the observed Fano factor with different modes' filling (Table 1):

| modes configuration | bulk $\boldsymbol{\nu_{\mathrm{b}}}$ | lower mode $\boldsymbol{\nu_{\mathrm{int}}^{l}=\nu_{\mathrm{b}}-\nu_{\mathrm{g}}^{l}}$ | upper mode $\boldsymbol{\nu_{\mathrm{int}}^{u}=\nu_{\mathrm{b}}-\nu_{\mathrm{g}}^{u}}$ | transmission ($t$) & Fano factor ($\boldsymbol{F}$) |
|---|---|---|---|---|
| *integer modes* | **1** | **1** | **1** | $t\sim 60\%$: $\boldsymbol{F=0.97}$ (ED Fig. 2a) |
| *integer & fractional* | **1** | **1** | **1/3** | $t\sim 82\%$: $\boldsymbol{F=0.93}$ (Fig. 2d, blue plots). |
| *fractional modes symmetric* | **1** | **1/3** | **1/3** | $t\sim 88\%$: $\boldsymbol{F=0.97}$ (Fig. 2d, red plots). |
| *fractional modes asymmetric* | **1** | **1/3** | **2/5** | $t\sim 91\%$: $\boldsymbol{F=1.0}$ (ED Fig. 2b). |
| *asymmetric* | **1** | **1/3** | **not quantized** | $t\sim 90\%$: $\boldsymbol{F=1.0}$ (ED Fig. 2b). |

**Table 1.**

In all the cases ***F*~1**. Note that uncertainty of the exact electron temperature and the slight non-linearity of the transmission lead to the deviations from ***F*=1**.

***Fixed*** $\boldsymbol{\nu_{\mathrm{int}}^{l}=2/3}$**:** Two types of the 'electron-hole conjugated' interfaced modes, $\nu_{\mathrm{int}}^{l}=1_{\mathrm{b}}-1/3_{\mathrm{g}}=2/3_{\mathrm{int}}$ and $2/3_{\mathrm{b}}-0_{\mathrm{g}}=2/3_{\mathrm{int}}$ are tested. Edge reconstruction leads to two DS copropagating 1/3 modes joined by two upstream neutral modes (see ED Section-VII)[22,27]. An intermediate conductance plateau is observed within the QPC at $t$=0.5 (see ED Fig. 4). The fully transmitted outer 1/3 mode and the fully reflected inner 1/3 mode do not lead to shot noise. However, the observed noise on the $t$=0.5 plateau, resulting from fractionalization of the neutral modes, leads to a Fano factor that is equal to the bulk filling factor (Table 2).

| bulk configuration | bulk $\nu_b$ | lower mode $\nu^l_{int} = \nu_b - \nu^l_g$ | upper mode $\nu^u_{int} = \nu_b - \nu^u_g$ | transmission ($t$) & Fano factor ($F$) |
|---|---|---|---|---|
| *integer bulk* | **1** | **2/3** | **1/3** | $t$ = 50%: $\boldsymbol{F = 1.0}$ (Fig. 3a, red plots). |
| *fractional bulk* | **2/3** | **2/3** | **1/3** | $t$ = 50%: $\boldsymbol{F = 0.64}$ (Fig. 3a, blue plots). |

**Table 2.**

The Fano factor remains close to the bulk filling for a wide range of QPC transmissions; i.e., away from the conductance plateau (Fig. 3b). When the QPC is strongly pinched, quasiparticle bunching[34] leads to $F$~1 at lower bias in all cases (see ED Fig. 5). See ED Section-V for more data of fractional edge modes, $2/3_b$ - $1/3_g$= $1/3_{int}$, $2/3_b - 4/15_g = 2/5_{int}$, with $\boldsymbol{F \cong \nu_b}$.

## IV. Theoretical model & Discussion

Our theoretical paradigm relies on the ubiquitous presence of counter-propagating neutral modes in the lowest spinless Landau level, being topological (for particle-hole conjugated states) or born due to spontaneous edge-reconstruction (for integer and particle-like states)[6,7,23,26], prone when the confining potential at the physical edge is not perfectly sharp[35,36].

We chose to consider here the simplest example of a reconstructed $\nu$ =1/3 bulk filling[7] in the orthodox QPC geometry, as this treatment captures the main physics of the universality of the quantum shot noise. In a general case, we assume the formation of a single $1/m$ fractional strip at the edge of the 2D bulk, with two counter-propagating $1/m$ modes (Figs. 4a & 4b). Accounting for inter-mode interaction and tunneling, the modes structure, shown in Fig. 4b, renormalizes to two DS modes with charge $(1/3 - 1/m)e$ and $e/m$ (from the inner to the outer), and a US neutral mode (Fig. 4c)[22,28]. In Fig. 4d we plot the DS ‘hot’ charge modes that emanate from the biased source S1 (solid lines), while the DS ‘cold’ emanating from the grounded S2 (broken lines). Source S1 injects N quasiparticles in each mode in a time interval $\tau$ (total current from source $I_{dc}$ =

$Ne/3\tau$). With the QPC tuned to partition a fraction $f$ of the outer most $1/m$ charge mode, its total transmission is $t = 3(1-f)/m$.

On the way to D1 and D2 the 'hot' and 'cold' modes equilibrate to the same chemical potential. The equilibration process keeps the currents arriving at the respective drains unchanged. Following this equilibration, the number of quasiparticles in each mode is: $N_1 = (1-f)3N/m$ near D1 and $N_2 = [m - 3(1-f)]3N/m$ near D2 (Fig. 4d). These equilibration process releases energy, and in order to conserve the appropriate quantum numbers[14,28,37], they lead to the creation of *neutralon* excitations (quasiparticles of the neutral modes characterized by non-trivial statistics). The latter are transported US from the vicinity of D2 and D1 in the direction of the QPC (and the sources). Along their trajectories, they fully decay into stochastic particle-hole pairs in the adjacent charge modes (as random pulses, equally probable particles and holes)[26]. The excitations in these modes near S1 and near S2 are schematically shown in Fig. 4e. Considering a fraction $f$ of tunneling events between the inner modes, the total charge reaching D1 is $Q_{\mathrm{D1}}$ (Fig. 4e & ED Section-IX) with the average $\overline{Q_{\mathrm{D1}}} = \frac{(1-f)3Ne}{m}$. The charge fluctuations, given by the autocorrelation of $Q_{\mathrm{D1}}$ measured over time $\tau$ at D1, equal $\frac{(1-f)(m-3)Ne^2}{3m^2}$. Together with the orthodox beam partitioning contribution[30,33] of $\frac{f(1-f)Ne^2}{m^2}$, the resulting total irreducible zero frequency current-current autocorrelation at D1 is $\ll I_{\mathrm{D1}} I_{\mathrm{D1}} \gg_{\omega=0} = \frac{2(1-f)(m+3f-3)Ne^2}{3m^2\tau}$. Expressing $f$ in terms of $t$, the effective Fano factor is then given by $F = \frac{\ll I_{\mathrm{D1}} I_{\mathrm{D1}} \gg_{\omega=0}}{2I_{\mathrm{dc}} \times e \times t(1-t)} = \frac{1}{3}$. Here, the Fano factor represents the bulk filling as well as the quasiparticle charge.

The latter statement does not hold for other particle-like states, e.g. $\nu$ =2/5 $\nu$ =3/7, etc., or even for particle-hole conjugated states, e.g. $\nu$ =2/3, $\nu$ =3/5, $\nu$ =4/7, etc. In the ED Section-IX we present a general calculation for an arbitrary filling with edge-reconstruction and the ensuing renormalization leading to the emergence of neutral modes. Furthermore, we present a similar analysis for less conventional geometry with an engineered interfaced boundary. Remarkably, in all these cases the Fano factor coincides with the bulk filling factor. We note that the Fano factor is unaffected by the partitioning of the charged quasiparticles (represented by the factor $f$, see Figs. 4d & 4e). Hence, it retains its value for $f = 0$, i.e., on an intermediate conductance plateau within the QPC[26,27]. We stress that details of the edge reconstruction are redundant as long as the filling fraction of the reconstructed side strip, $1/m$, is finite (see Fig. 4). Even minor reconstruction $1/m \ll 1$ leads to the universal result, $\boldsymbol{F}=\boldsymbol{\nu}_{\mathbf{b}}$. We note, though, that our results are compatible with the orthodox beam partitioning mechanism for shot noise, expected in the absence of edge reconstruction (when no neutral modes emerge).

At higher temperatures, the neutral excitations are known to decay[21], and one expects the (off-the-plateau) Fano factor to represent the quasiparticle charge[4,21]. Determining the quasiparticle charge at low temperatures is crucial. A simple path for extracting this charge combines experiment and theory. One measures the noise on and off the conductance plateau, calculates the neutral mode contribution to the latter, and then extracts the beam-partitioning contribution to the noise off the plateau. This last contribution is a marker of the quasi-particle charge. We note, though, that $1/m$ may be exceedingly small, resulting in a blurred plateau with an elusive conclusion concerning the quasi-particle charge and the on-the-plateau noise. Further protocols and geometries are needed to determine the partitioned charge accurately.

## Conclusions

Contrary to current understanding, shot noise in the fractional quantum Hall effect does not reveal the quasiparticle charge at sufficiently low temperatures. Our novel new measurements (consistent with previous results) show, systematically and reproducibly, that the Fano factor is identical to the fractional state filling of the bulk, $F = \nu_{\mathrm{b}}$. Here, we expanded the actual realizations by breaking 'bulk – edge' correspondence. In the new realization, the orthodox structure of the partitioning quantum point contact (QPC) constriction was replaced by a QPC that partitions a chiral interface mode at the junction between two different QHE states. In all the realizations, we provided a microscopic explanation of the results by considering the omnipresence of neutral modes. The interplay between charge and neutral modes provides a new platform to the generation of shot noise. Regardless of the presence or absence of an additional conventional channel partitioned shot noise, these conspire to yield a Fano factor $\boldsymbol{F} = \boldsymbol{\nu}_{\mathbf{b}}$, a result that was far unexpected. There are quite a few future challenges, e.g., checking whether this universal behavior holds for non-abelian states like 5/2[20,33].

**References**

1 Laughlin, R. B. Anomalous quantum Hall effect - an incompressible quantum fluid with fractionally charged excitations. *Physical Review Letters* **50**, 1395-1398, doi:DOI 10.1103/PhysRevLett.50.1395 (1983).

2 dePicciotto, R. *et al.* Direct observation of a fractional charge. *Nature* **389**, 162-164, doi:Doi 10.1038/38241 (1997).

3 Reznikov, M., de Picciotto, R., Griffiths, T. G., Heiblum, M. & Umansky, V. Observation of quasiparticles with one-fifth of an electron's charge. *Nature* **399**, 238-241, doi:Doi 10.1038/20384 (1999).

4 Chung, Y. C., Heiblum, M. & Umansky, V. Scattering of bunched fractionally charged quasiparticles. *Physical Review Letters* **91**, 216804, doi:10.1103/PhysRevLett.91.216804 (2003).

5 Bid, A., Ofek, N., Heiblum, M., Umansky, V. & Mahalu, D. Shot noise and charge at the 2/3 composite fractional quantum Hall state. *Physical Review Letters* **103**, 236802, doi:10.1103/PhysRevLett.103.236802 (2009).

6 Bhattacharyya, R., Banerjee, M., Heiblum, M., Mahalu, D. & Umansky, V. Melting of interference in the fractional quantum Hall effect: Appearance of neutral modes. *Physical Review Letters* **122**, 246801, doi:10.1103/PhysRevLett.122.246801 (2019).

7 Inoue, H. *et al.* Proliferation of neutral modes in fractional quantum Hall states. *Nature Communications* **5**, 4067, doi:10.1038/ncomms5067 (2014).

8 Prange, R. & Girvin, S. M. *The Quantum Hall Effect*. (Springer Verlag, 1990).

9 Sarma, S. D. & Pinczuk, A. *Perspective in Quantum Hall Effects*. (Wiley, 1996).

10 Halperin, B. I. & Jain, J. K. *Fractional Quantum Hall Effects: New Developments*. (WORLD SCIENTIFIC, 2020).

11 Halperin, B. I. Quantized Hall conductance, current-carrying edge states, and the existence of extended states in a two-dimensional disordered potential. *Physical Review B* **25**, 2185-2190, doi:DOI 10.1103/PhysRevB.25.2185 (1982).

12 Wen, X.-G. *Quantum field theory of many-body systems: From the origin of sound to an origin of light and electrons.* (Oxford University Press on Demand 2004).

13 Kane, C. L. & Fisher, M. P. A. in *Perspectives in Quantum Hall Effects: Novel Quantum Liquids in Low-Dimensional Semiconductor Structures* (eds S. Das Sarma & A. Pinczuk) (John Wiley, 1996).

14 Wen, X. G. Gapless boundary excitations in the quantum Hall states and in the chiral spin states. *Physical Review B* **43**, 11025-11036, doi:10.1103/PhysRevB.43.11025 (1991).

15 Stern, A. Anyons and the quantum Hall effect - A pedagogical review. *Annals of Physics* **323**, 204-249, doi:10.1016/j.aop.2005.03.001 (2008).

16 Zheng, H. Z., Wei, H. P., Tsui, D. C. & Weimann, G. Gate-controlled transport in narrow $GaAs/Al_xGa_{1-x}As$ heterostructures. *Physical Review B* **34**, 5635 (1986).

17 Heiblum, M. & Feldman, D. E. Edge probes of topological order. *International Journal of Modern Physics A* **35**, doi:10.1142/s0217751x20300094 (2020).

18 Chang, A. M. Chiral Luttinger liquids at the fractional quantum Hall edge. *Reviews of Modern Physics* **75**, 1449-1505, doi:DOI 10.1103/RevModPhys.75.1449 (2003).

19 Saminadayar, L., Glattli, D. C., Jin, Y. & Etienne, B. Observation of the e/3 fractionally charged Laughlin quasiparticle. *Physical Review Letters* **79**, 2526-2529, doi:DOI 10.1103/PhysRevLett.79.2526 (1997).

20 Dolev, M., Heiblum, M., Umansky, V., Stern, A. & Mahalu, D. Observation of a quarter of an electron charge at the $\nu$ = 5/2 quantum Hall state. *Nature* **452**, 829-834, doi:10.1038/nature06855 (2008).

21 Bid, A. *et al.* Observation of neutral modes in the fractional quantum Hall regime. *Nature* **466**, 585-590, doi:10.1038/nature09277 (2010).

22 Wang, J., Meir, Y. & Gefen, Y. Edge reconstruction in the $\nu$=2/3 fractional quantum Hall state. *Physical Review Letters* **111**, 246803, doi:10.1103/PhysRevLett.111.246803 (2013).

23 Khanna, U., Goldstein, M. & Gefen, Y. Emergence of Neutral Modes in Laughlin-like Fractional Quantum Hall Phases. *arXiv:2109.15293* (2021).

24 Meir, Y. Composite edge states in the $\nu$ = 2/3 fractional quantum Hall regime. *Physical Review Letters* **72**, 2624-2627, doi:10.1103/PhysRevLett.72.2624 (1994).

25 Hu, L. & Zhu, W. Abelian origin of $\nu$=2/3 and 2+2/3 fractional quantum Hall effect. *arXiv:2109.00781*.

26 Khanna, U., Goldstein, M. & Gefen, Y. Fractional edge reconstruction in integer quantum Hall phases. *Physical Review B* **103**, L121302, doi:10.1103/PhysRevB.103.L121302 (2021).

27 Sabo, R. *et al.* Edge reconstruction in fractional quantum Hall states. *Nature Physics* **13**, 491-496, doi:10.1038/Nphys4010 (2017).

28 Kane, C. L., Fisher, M. P. & Polchinski, J. Randomness at the edge: Theory of quantum Hall transport at filling *ν* = 2/3. *Physical Review Letters* **72**, 4129-4132, doi:10.1103/PhysRevLett.72.4129 (1994).

29 Lin, C. *et al.* Charge equilibration in integer and fractional quantum Hall edge channels in a generalized Hall-bar device. *Physical Review B* **99**, 195304, doi:10.1103/PhysRevB.99.195304 (2019).

30 Martin, T. & Landauer, R. Wave-packet approach to noise in multichannel mesoscopic systems. *Physical Review B* **45**, 1742-1755, doi:10.1103/PhysRevB.45.1742 (1992).

31 Buttiker, M. Scattering theory of current and intensity noise correlations in conductors and wave guides. *Physical Review B* **46**, 12485-12507, doi:10.1103/PhysRevB.46.12485 (1992).

32 Feldman, D. E. & Heiblum, M. Why a noninteracting model works for shot noise in fractional charge experiments. *Physical Review B* **95**, 115308, doi:10.1103/PhysRevB.95.115308 (2017).

33 Dolev, M. *et al.* Dependence of the tunneling quasiparticle charge determined via shot noise measurements on the tunneling barrier and energetics. *Physical Review B* **81**, 161303, doi:10.1103/PhysRevB.81.161303 (2010).

34 Comforti, E., Chung, Y. C., Heiblum, M., Umansky, V. & Mahalu, D. Bunching of fractionally charged quasiparticles tunnelling through high-potential barriers. *Nature* **416**, 515-518, doi:10.1038/416515a (2002).

35 Chamon, C. & Wen, X. G. Sharp and smooth boundaries of quantum Hall liquids. *Physical Review B* **49**, 8227-8241, doi:DOI 10.1103/PhysRevB.49.8227 (1994).

36 Khanna, U., Murthy, G., Rao, S. & Gefen, Y. Spin Mode Switching at the Edge of a Quantum Hall System. *Physical Review Letters* **119**, 186804, doi:10.1103/PhysRevLett.119.186804 (2017).

37 Park, J., Rosenow, B. & Gefen, Y. Symmetry-related transport on a fractional quantum Hall edge. *Physical Review Research* **3**, 023083, doi:10.1103/PhysRevResearch.3.023083 (2021).

## Acknowledgements

We acknowledge the continuous support of the Sub-Micron Center staff. M.H. acknowledges the support of the European Research Council under the European Community's Seventh Framework Program (FP7/2007-2013)/ERC under grant agreement number 713351, the partial support of the Minerva Foundation with funding from the Federal German Ministry for Education and Research, under grant number 713534. M.G. was supported by the Israel Science Foundation (ISF) and the Directorate for Defense Research and Development (DDR&D) grant No. 3427/21 and by the US-Israel Binational Science Foundation (BSF) Grants No. 2016224 and 2020072. Y.G. was supported by CRC 183 (project C01), the Minerva Foundation, DFG Grant No. RO 2247/11-1, MI 658/10-2, the German Israeli Foundation (Grant No. I-118-303.1-2018), the National Science Foundation through award DMR- 2037654 and the US-Israel Binational Science Foundation (BSF), and the Helmholtz International Fellow Award. A.D. was supported by the German-Israeli Foundation Grant No. I-1505-303.10/2019 and the GIF. A.D. also thanks Koshland Foundation for Koshland Fellowship, Israel planning and budgeting committee (PBC) and Weizmann Institute of Science, Israel Dean of Faculty fellowship for financial support.

## Author contributions

S.B. and R.B. fabricated the devices, performed the measurements, and analyzed the data with H.K.K. M.H. supervised the experiment and the analysis. Y.G. has contributed in conceiving the theoretical model. A.D., M.G. and Y.G. developed the theoretical model. V.U. grew the GaAs heterostructures. All authors contributed to writing of the manuscript.

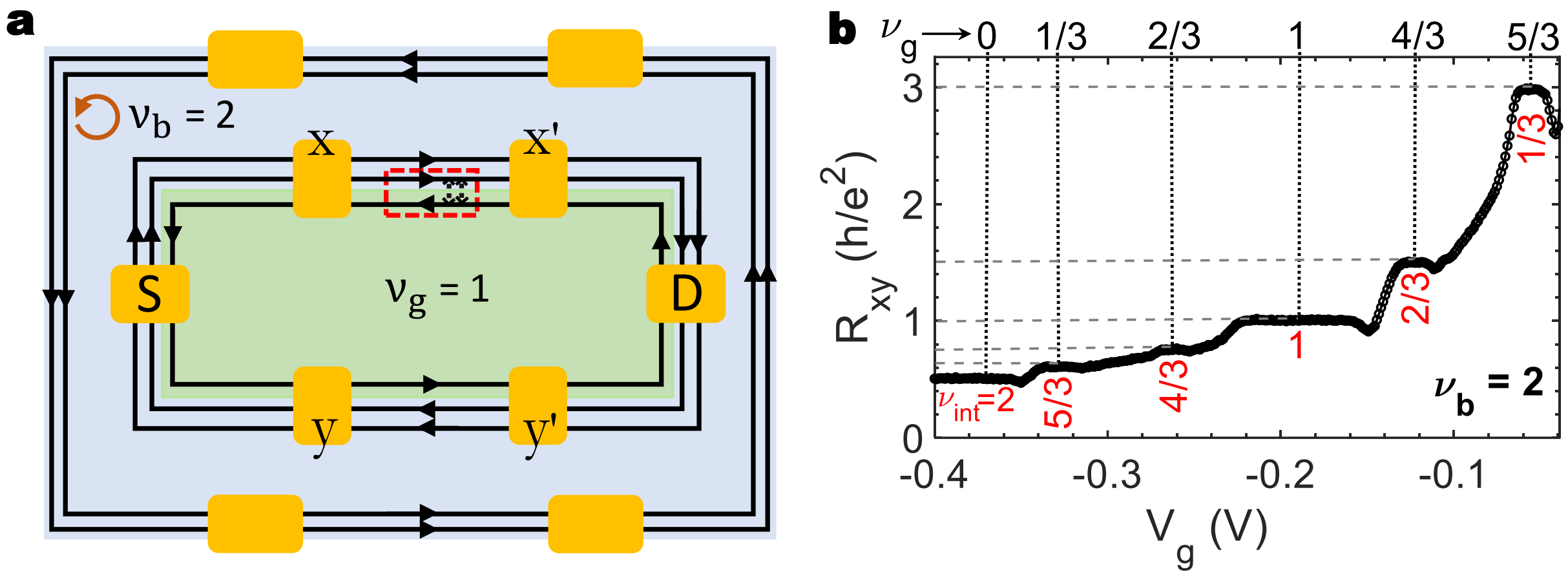

**Figure 1 | Interface edge modes. a,** Schematic of the platform with top gated bulk with filling $\nu_g = 1$ state (green). The ungated surrounding region is at $\nu_b = 2$ (blue). Counter-propagating interface edge modes equilibrated with an 'effective filling' $\nu_{int} = \nu_b - \nu_g$. Source contact S and Drain contact D (yellow) are placed at the interface. X and Y contacts are used to measure the 4-probe Hall conductance, $\sigma_{xy} = \nu_{int}\frac{e^2}{h}$ of the interface edge. **b,** Interface conductance for $\nu_b = 2$ as function of center gate voltage (filling factor of center bulk), exhibiting integer and fractional plateaus of the interface modes.

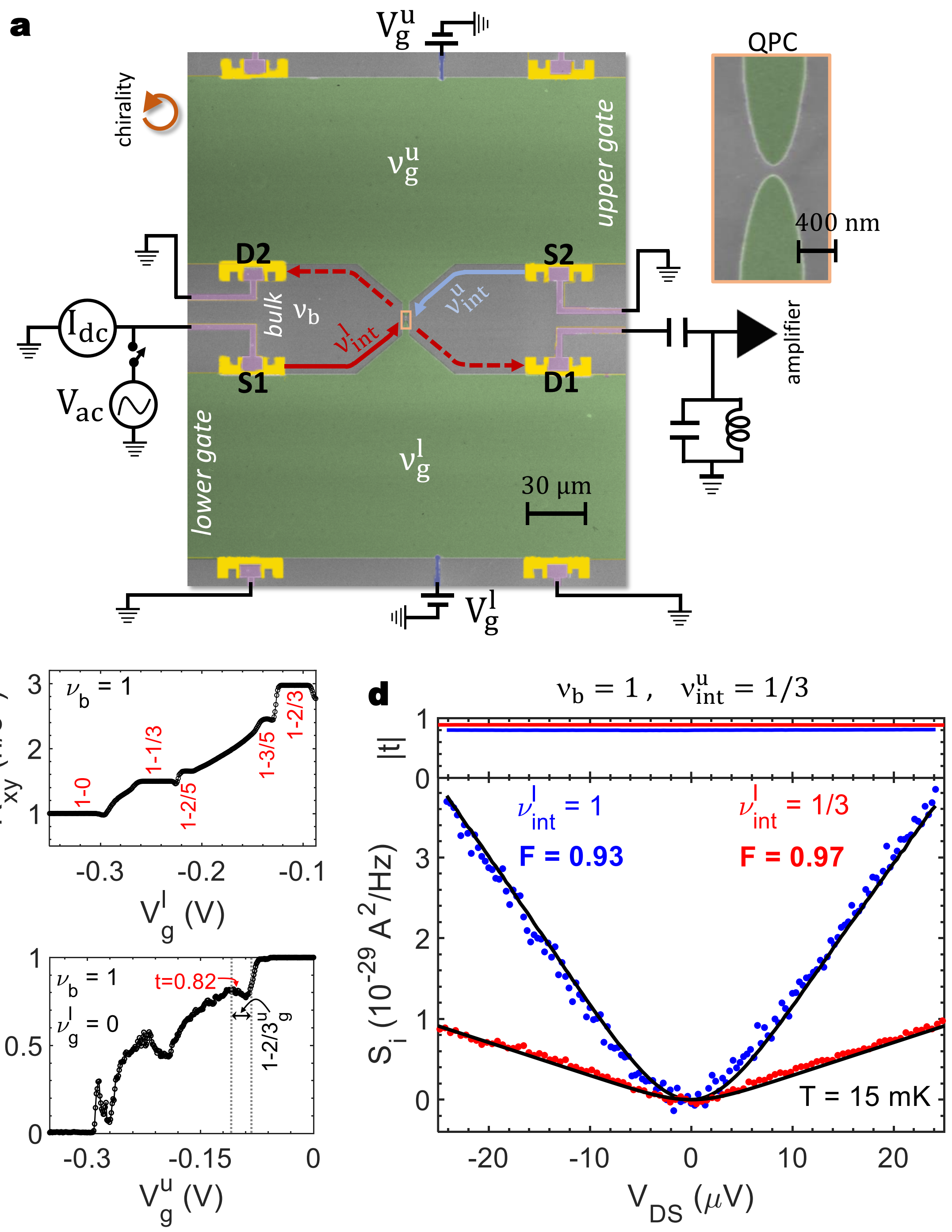

**Figure 2 | Experimental setup for shot noise measurement with interface modes. a,** False color SEM image of the 'QPC geometry' formed by two peninsulas of two separate gates. Inset: image of the QPC. The ohmic contacts (yellow) were made by alloying Au/Ge/Ni, and gates (green) were formed by evaporating a thin PdAu/Au film on top of a 25 nm atomic-layer-deposited $HfO_2$. Metallic air-bridges (dark blue) are connected to the gates for applying gate voltage. Source contact S1 (connected to the DC bias), and Drain contact D1 (connected to the amplifier), are placed at

the lower gate interface. A DC voltage $V_g^l$ sets the filling factor underneath, thereby the incoming interface mode. The gate voltage $V_g^u$ on the upper gate tunes the QPC back-scattering, but does not guarantee the filling underneath to be the same as in the lower gate. Hot (biased) and cold (grounded) edge modes are shown in red and light-blue lines, respectively. The source - QPC distance is ~50 μm, while charge equilibration at the interface is established within a few microns. **b**, Four-terminal conductance measurements as a function of $V_g^l$ showing the formation of fractional edge modes around the lower gate at $\nu_b$=1. **c,** QPC transmission *t* as function of $V_g^u$, while $V_g^l$ is fixed such that $\nu_g^l$=0; i.e., the incoming edge mode is 1-0. **d**, Measured Fano factors for 1-0=1 and 1-2/3=1/3 interface modes. Top panel is bias dependent QPC transmission; t=0.82 for $1_b$-$0_g$ (blue) and t=0.88 for $1_b$-$2/3_g$ (red). Bottom panel is the measured spectral density of DS current fluctuation with $V_{DS}$ (i.e., $I_{dc} \times R_{xy}$), shown in dots. Black solid line is the fit. For both $1_b$-$0_g$=$1_{int}$ (blue) and $1_b$-$2/3_g$=$1/3_{int}$ (red), Fano factor $F \sim 1$. Estimated Fano factors are within an accuracy of $\pm 0.05$. Obtained electron temperature is 15mK.

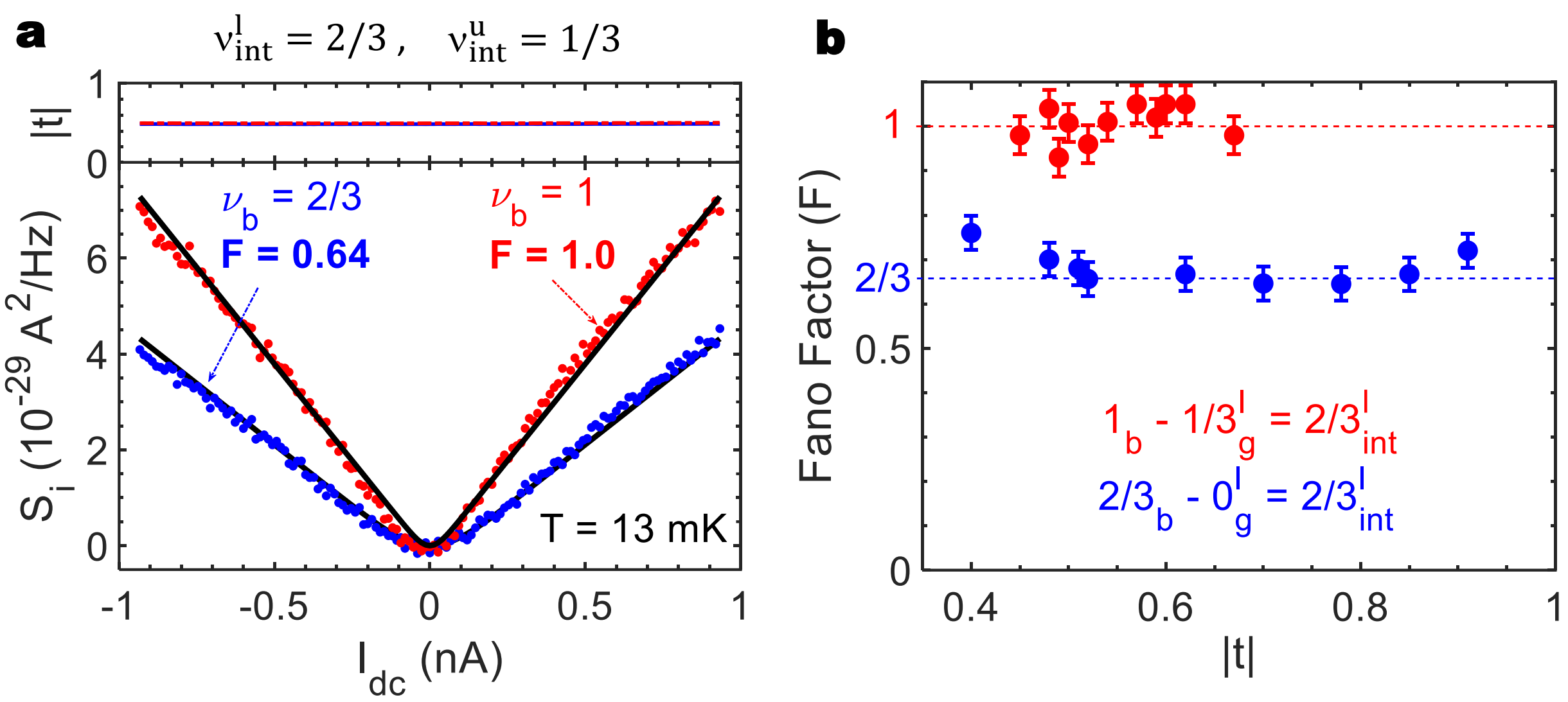

**Figure 3 | Shot noise with two types of interface 2/3 modes: 1-1/3 and 2/3-0. a,** Shot noise measurements at $1_b$-$1/3_g$ and $2/3_b$-$0_g$ on the plateau of $t$=1/2. **Top panel** - Bias dependence of the transmission. **Bottom panel** – excess noise data at $1_b$-$1/3_g$ (red dots) and at $2/3_b$-$0_g$ (blue dots) with $I_{dc}$. Black solid lines are the fits, with the appropriate Fano factors. **c,** Fano factor = bulk filling, measured over a wide range of QPC transmission. The accuracy of the estimated Fano factor is within ±0.04 for both the cases.

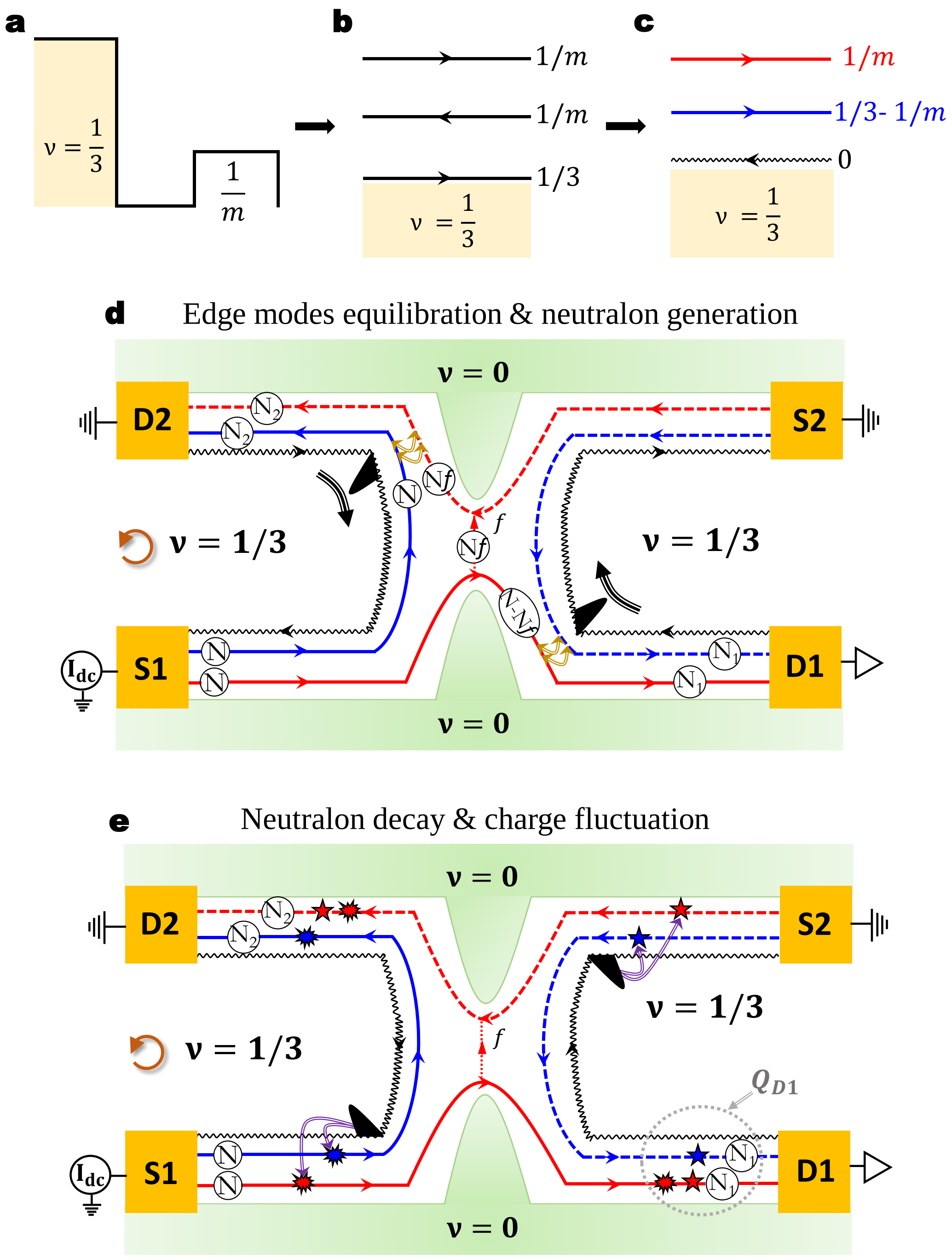

**Figure 4 | Theoretical model with the example of reconstructed ν=1/3. a,** The electronic occupation near the edge for ν=1/3 with a *1/m* reconstruction. **b,** The resulting bare edge modes. **c,** renormalized edge: the outer mode remains disconnected, while the two inner modes are renormalized, giving rise to a redefined DS charge mode and a counter-propagating neutral mode. The characterization of the modes, from inner to outer, is 1/3-*1/m* (blue), and *1/m* (red). **d,** The QPC with bulk filling 1/3, following *1/m* edge reconstruction and renormalization. *f* is the

tunneling probability between the two *1/m* charge modes, resulting in charge transmitted to D1 (red). The number of charge particles (measured over a time interval $\tau$ ) in the respective modes at different stages along the propagation is schematically shown in small circles. Equilibration between charge modes is shown by dark-yellowish double-arrow; the black bumps represent excitations (neutralons) in the neutral mode, which propagate US. **e,** Decay (shown by purple arrows) of neutralons, leading to the stochastic generation of charge (quasiparticle/quasihole) pulses near S1 (represented by 'beetles') and near S2 (represented by 'stars'). All charge contributions reaching the drain D1 (and the amplifier) are depicted in the big-dotted circle (total charge $Q_{\mathrm{D1}}$). For a quantitative analysis, see ED Section-IX.

# Extended Data

Extended shot-noise results with different configurations at integer and fractional bulk filling factor, US neutral noise measurements and detailed theoretical analysis.

## I. Shot-noise gain calibration:

To calibrate the gain of the cold amplifier (CA), we measure the shot-noise by weak-partitioning of the outer edge of filling factor (2-0), see ED Fig. 1a. Fit to the data with F = 1 determines the gain $G_0$ of the amplifier and the electron temperature. The calibration is further verified with similar gain obtained at ν = 2 outer edge using a regular QPC, see ED Fig. 1b. However, this is an effective gain which depends on the bandwidth of the LCR circuit. This effective gain is related to the area of the power response of the LCR circuit. It is easily derived that the square root of the area ratio in two filling factors is equal to the ratio of their effective gain. Thus, the gain $G_1$ at another filling factor is calculated, and used in the fitting equation to estimate the free parameter, the Fano factor (*F*).

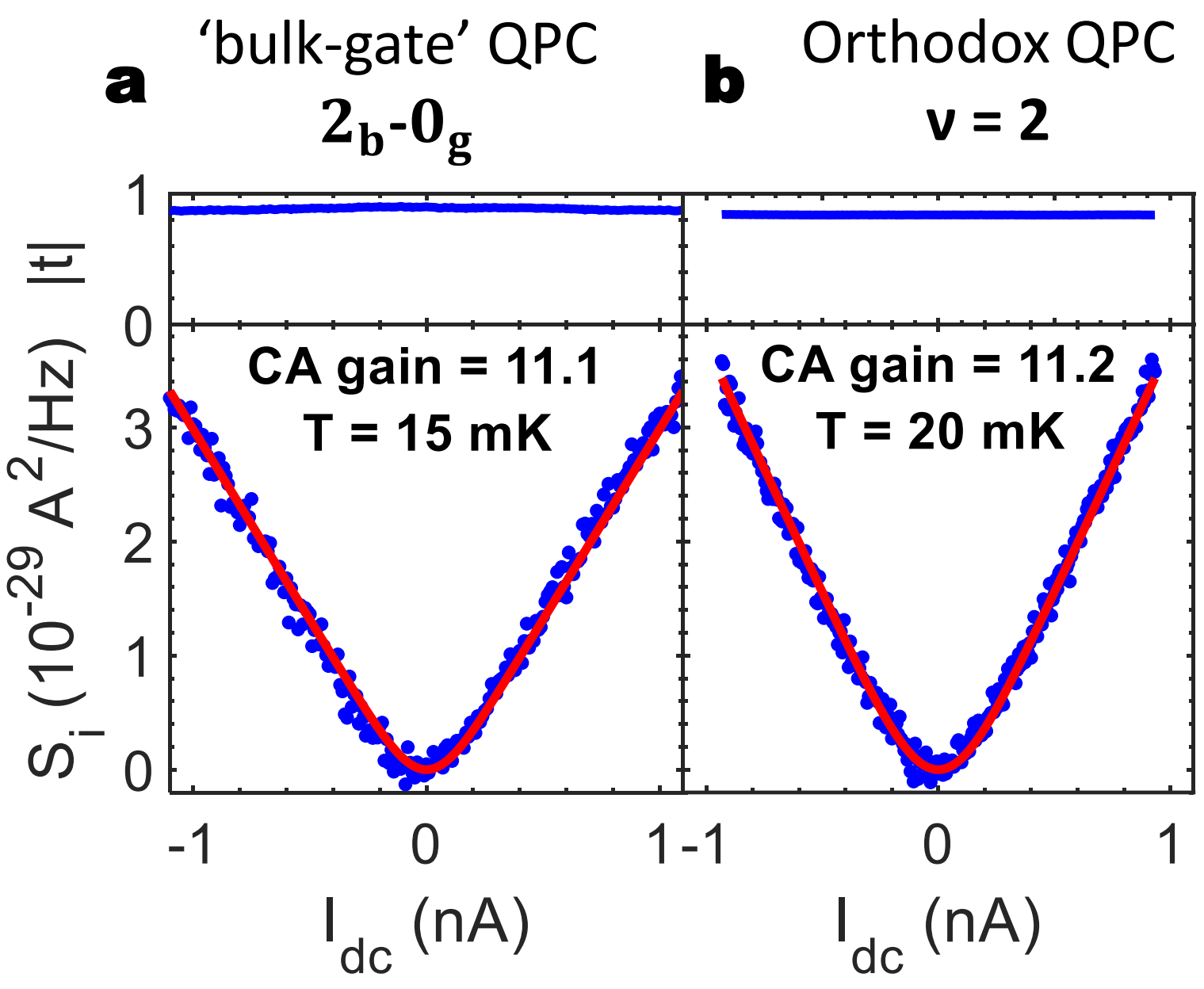

**ED Figure 1 | Cold amplifier (CA) gain calibration. a,** Shot-noise at 2-0 outer edge in 'bulk-gate' QPC geometry. The fit assuming *F*=1 gives the gain of 11.1. **b,** Same measurement in a regular QPC and the obtained gain is 11.2.

## II. Shot noise at bulk $\nu_b = 1$ with symmetric and asymmetric QPC

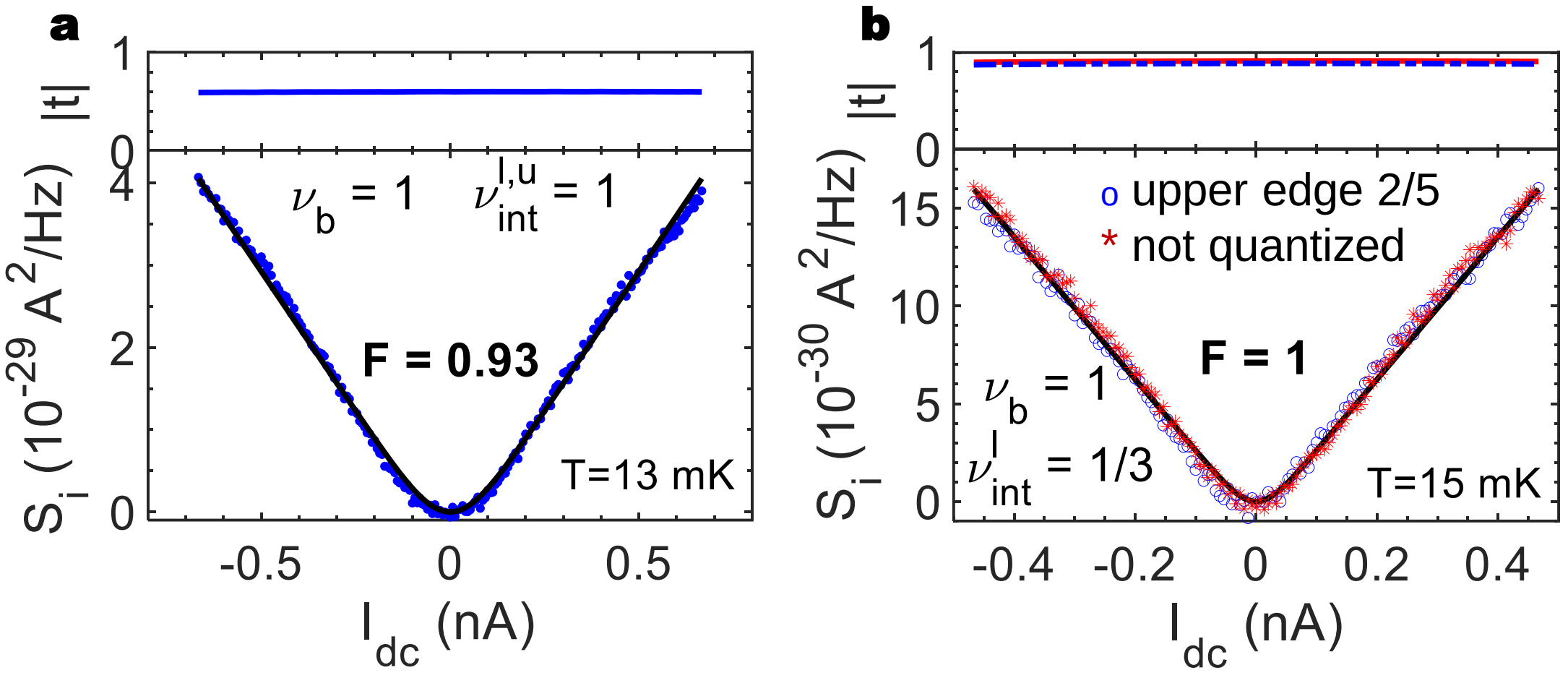

**ED Figure 2 | Shot noise at integer bulk $\nu_b = 1$. a,** DS current noise ($S_i$) at bulk filling factor 1 for the simplest conventional case, i.e., both modes are also 1. Fano factor ~1. **b,** Shot noise data for partitioning $1_b$-$2/3_g$=$1/3_{int}$ mode when the QPC is asymmetric with $\nu^l_g \neq \nu^u_g$ and $\nu^u_g$ not quantized. We obtained $F = 1$, with accuracy $\pm 0.05$. The electron temperature obtained is slightly higher (~3-4 mK) than the fridge's base temperature.

## III. Longitudinal resistance measurement at bulk $\nu_b = 1$

Four-terminal longitudinal resistance ($R_{xx}$) measurement along the interface at bulk filling $\nu_b = 1$. $R_{xx}$ for $1_b$-$0_g$=$1_{int}$, $1_b$-$1/3_g$=$2/3_{int}$, and $1_b$-$2/3_g$=$1/3_{int}$ nearly vanish, showing the bulk to be insulating in both gated and ungated regions in the device.

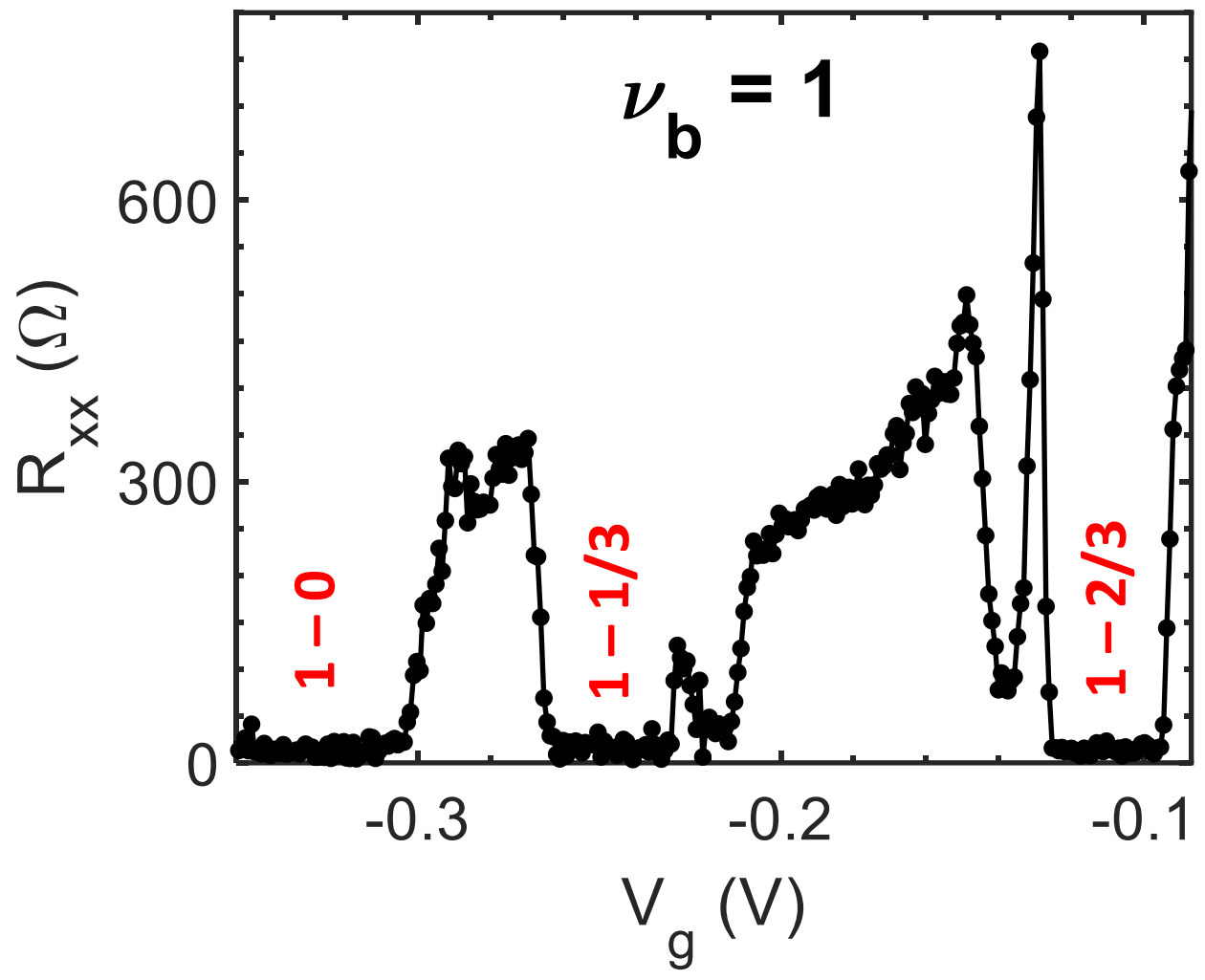

**ED Figure 3 |** Longitudinal interface resistance at $\nu_b$ =1.

## IV. e-h-conjugated 2/3 mode

### (a) QPC transmission

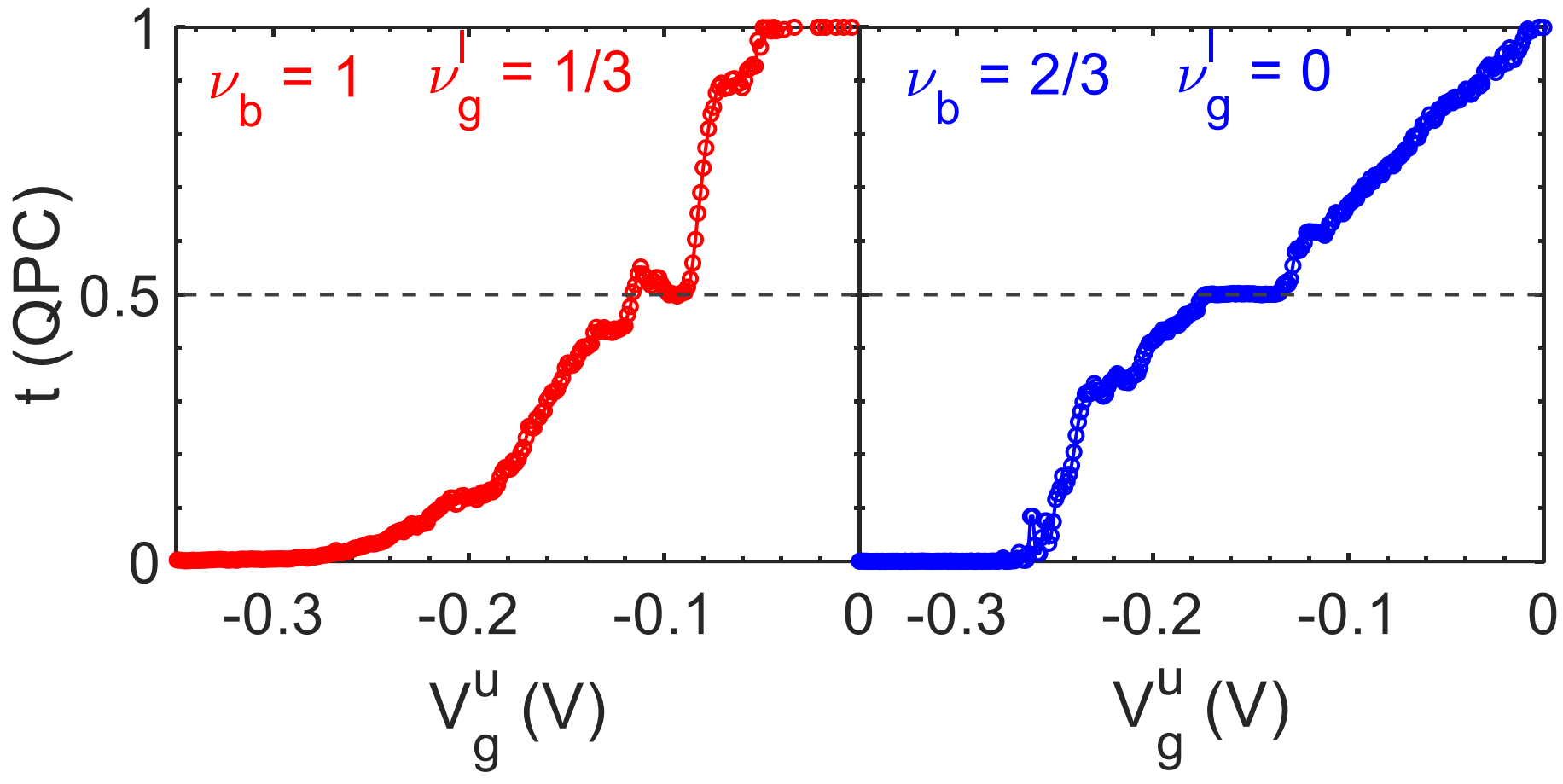

**ED Figure 4 | QPC responses for two types of interface 2/3 modes: 1-1/3 and 2/3-0.** QPC transmission of a 2/3 interface mode at the lower-gate-bulk interface as function of upper-gate voltage via interfacing $1_b$-$1/3_g$ (left) and $2/3_b$-$0_g$ (right). The transmission at the QPC with $V_g^u$ shows the common plateau at $t$=1/2 in both cases, indicating edge reconstruction.

**(b) Bunching phenomenon at low transmission (strong back-scattering)**

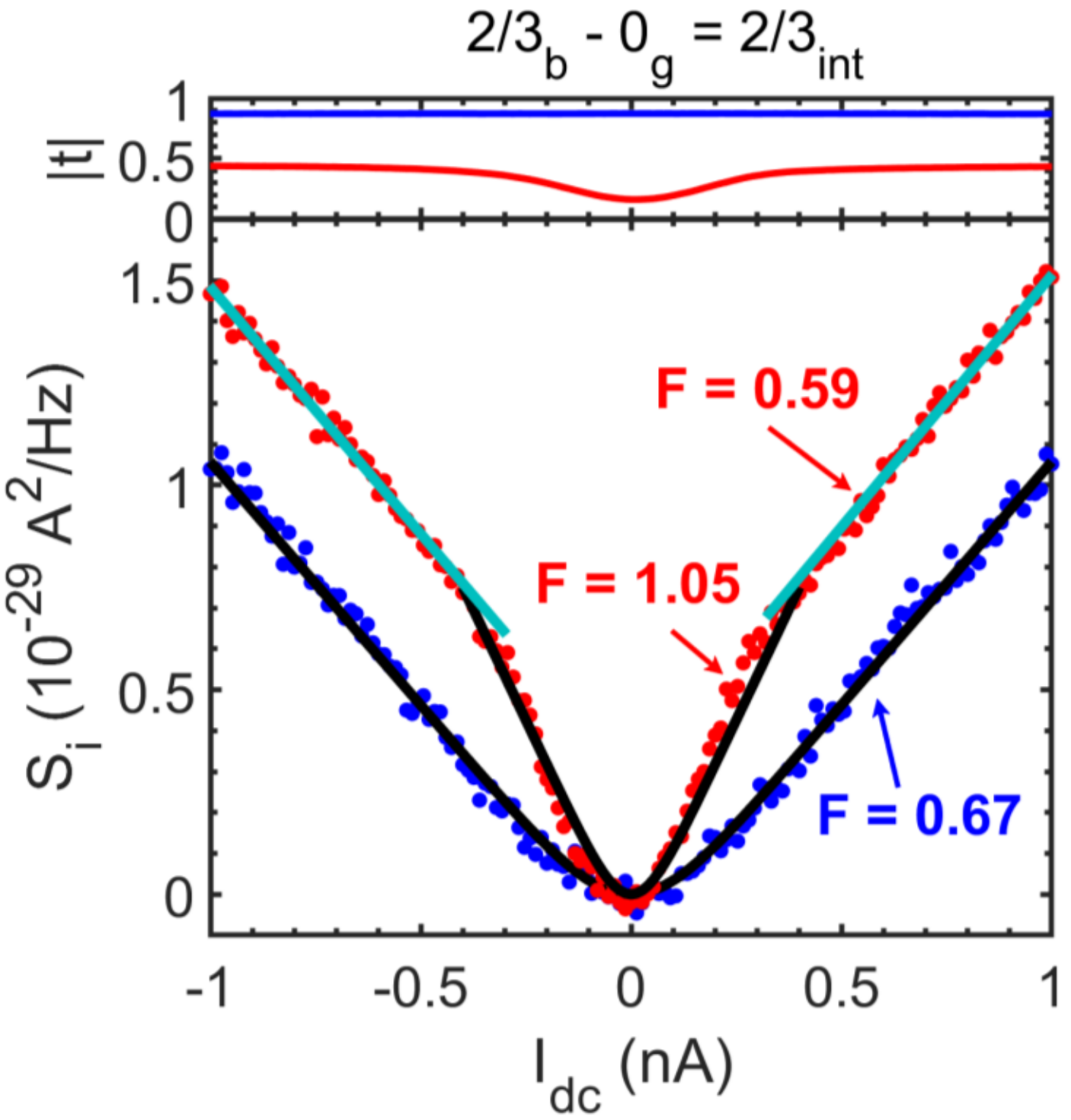

**ED Figure 5 | Bunching of 2/3-0 edge mode.** Quasiparticle bunching leads to electron tunneling at low transmission and small impinging bias current (red plots). At higher bias, the QPC transmission increases and Fano factor is close to 2/3. This is consistent with the result of 2/3 edge in a regular QPC[1]. For a comparison, the blue plots represent the data for a higher transmission with $F = 2/3$ over the full bias current range.

## V. Fractional states $\nu_{int} = 1/3$ and $\nu_{int} = 2/5$ at bulk $\nu_b = 2/3$:

Two fractional modes with $\frac{1}{3}e^2/h$ conductance and $\frac{2}{5}e^2/h$ conductance at the lower interface were engineered by interfacing $2/3_b$-$1/3_g$, and $2/3_b$-$4/15_g$, respectively (4/15 is a new quantum Hall state, which we can stabilize only by gating). Unlike ubiquitous 1/3 and 2/5 modes, the interface $1/3_{int}$ and $2/5_{int}$ modes are expected to have negative or zero thermal conductance ($K_{xy}$), hence carry topological neutral modes. As expected, finite US noise is observed (ED Figs. 6a & 6b). While measuring shot noise on these interface edge modes in a symmetric QPC, we found the Fano factors close to 2/3 (ED Figs. 6c & 6d). Similar results agreeing with $F = \nu_b$ are obtained for an asymmetric QPC (not shown here).

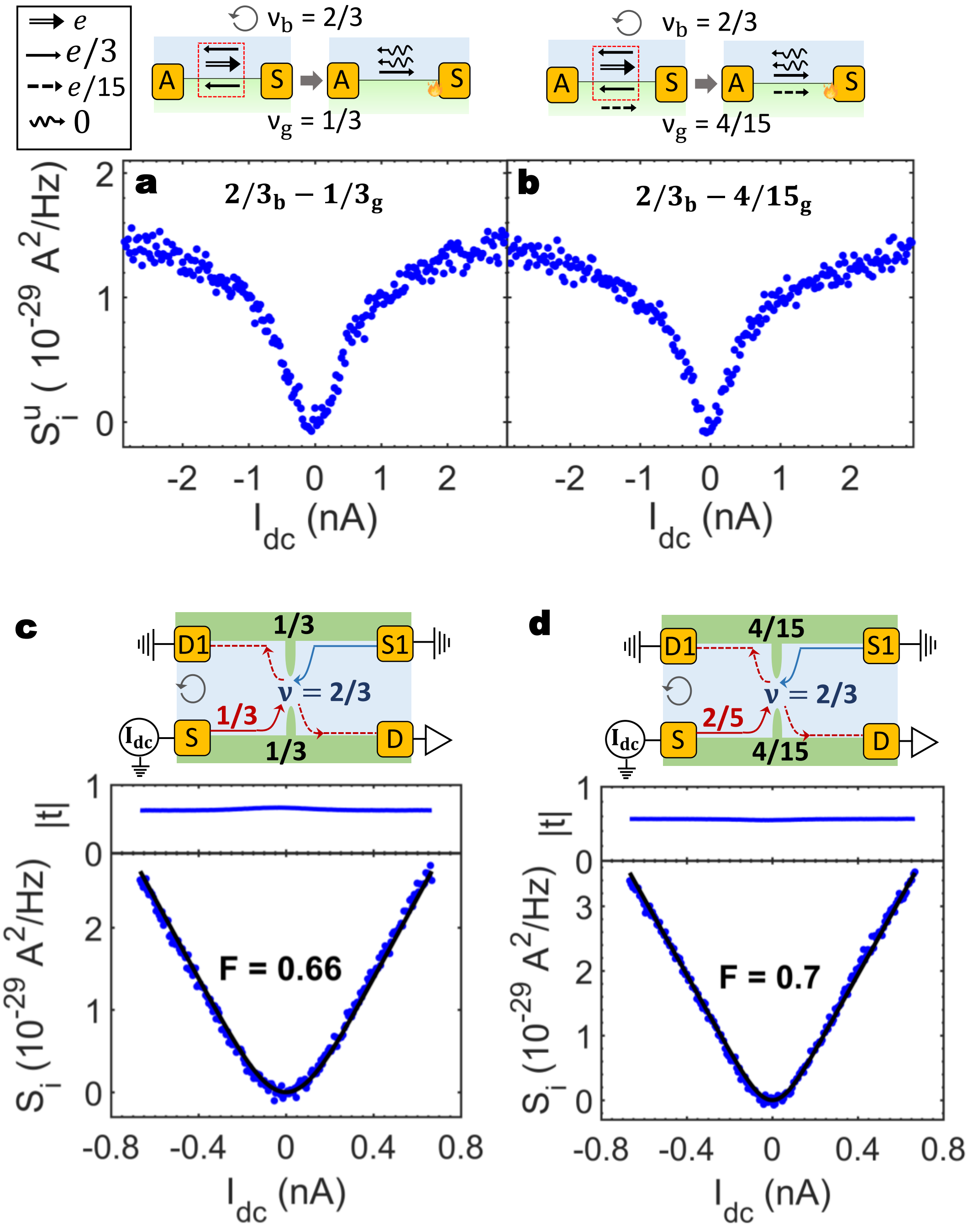

**ED Figure 6 | Fano factor and US noise of lowest fractional states. a,b,** Spectral density of US current noise due to topological neutral modes of unconventional particle-like $1/3_{int}$ and $2/5_{int}$ at $2/3_b$. **c,d,** Spectral density of DS current noise of 1/3 by interfacing $2/3_b$-$1/3_g$ for $t$=0.62 and of 2/5 by interfacing $2/3_b$-$4/15_g$ for $t$=0.56. Black dots are the measured data and blue solid lines are the fit. Estimated Fano factors are always close to 2/3. Schematics in the respective inset describe the interface mode incident on the QPC and the corresponding bulks.

## VI. Shot noise for $\nu_{int} = 1$ at $\nu_b = 2$

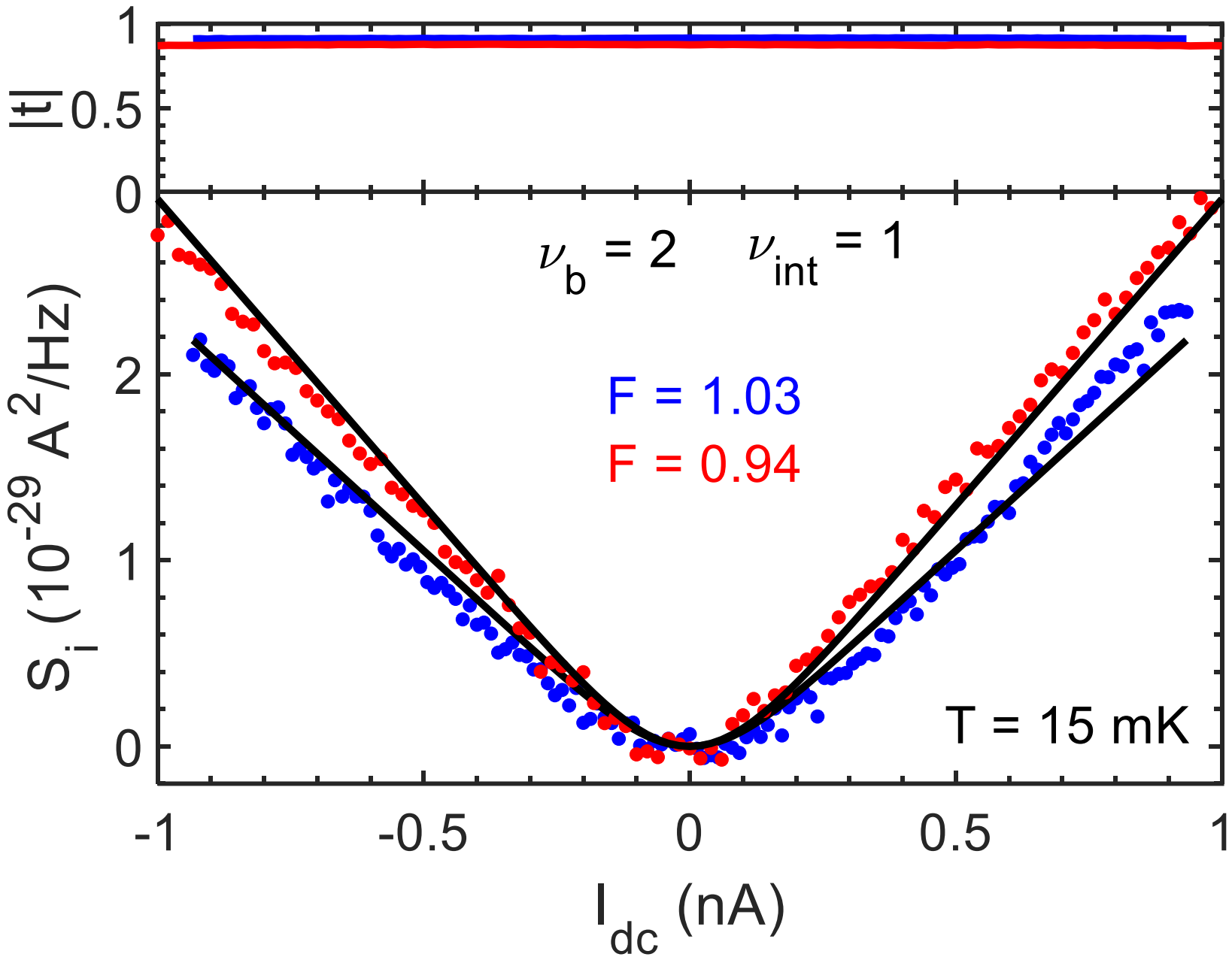

**ED Figure 7 |** Shot noise measurement on weak partitioning of $2_b$-$1_g$=$1_{int}$ edge mode, with $F \sim 1$. Data for two QPC transmissions (t=0.91 and t=0.87) are shown. This indicates that $F$ is limited to the value of single electron charge in the case of weak backscattering in a QPC in our measurement platform, as has been demonstrated in previously reported experiments with conventional geometries. Our theory predicts the same even if $\nu_b = 2$.

## VII. US neutral noise

ED Figure 8 shows the measured current noise for the interface edge modes (primarily discussed in the main text) in the US direction 70 μm away from the source hot-spot. $1_b$-$0_g$=$1_{int}$ and $1_b$-$2/3_g$=$1/3_{int}$ modes do not carry topological neutral mode and thus zero noise is measured at the US amplifier (ED Figs. 8a & 8b). However, short length non-topological neutral modes due to edge reconstruction proliferate[2,3].

The conventional edge mode $2/3_b$-$0_g$, supports a DS charge mode with conductance $\frac{2e^2}{3h}$ and an US neutral mode[4,5] with zero thermal conductance[6] (ED Fig. 8c). Similar US noise was measured at the interface $1_b$-$1/3_g$=$2/3_{int}$ (ED Fig. 8d). However, a novel 2/3 interface edge mode, without a topological neutral mode, can be engineered at the interface of $4/3_b$-$2/3_g$=$2/3_{int}$. This 2/3 mode supports only two DS charge modes, each with a conductance $\frac{e^2}{3h}$ (ED Fig. 8e). This is an enlightening case of the $\nu = 2/3$ state different from the usually known one. The thermal conductance expected for such a novel 2/3 state is 2 quanta of $K_{xy}$, which is unique and yet to be measured.

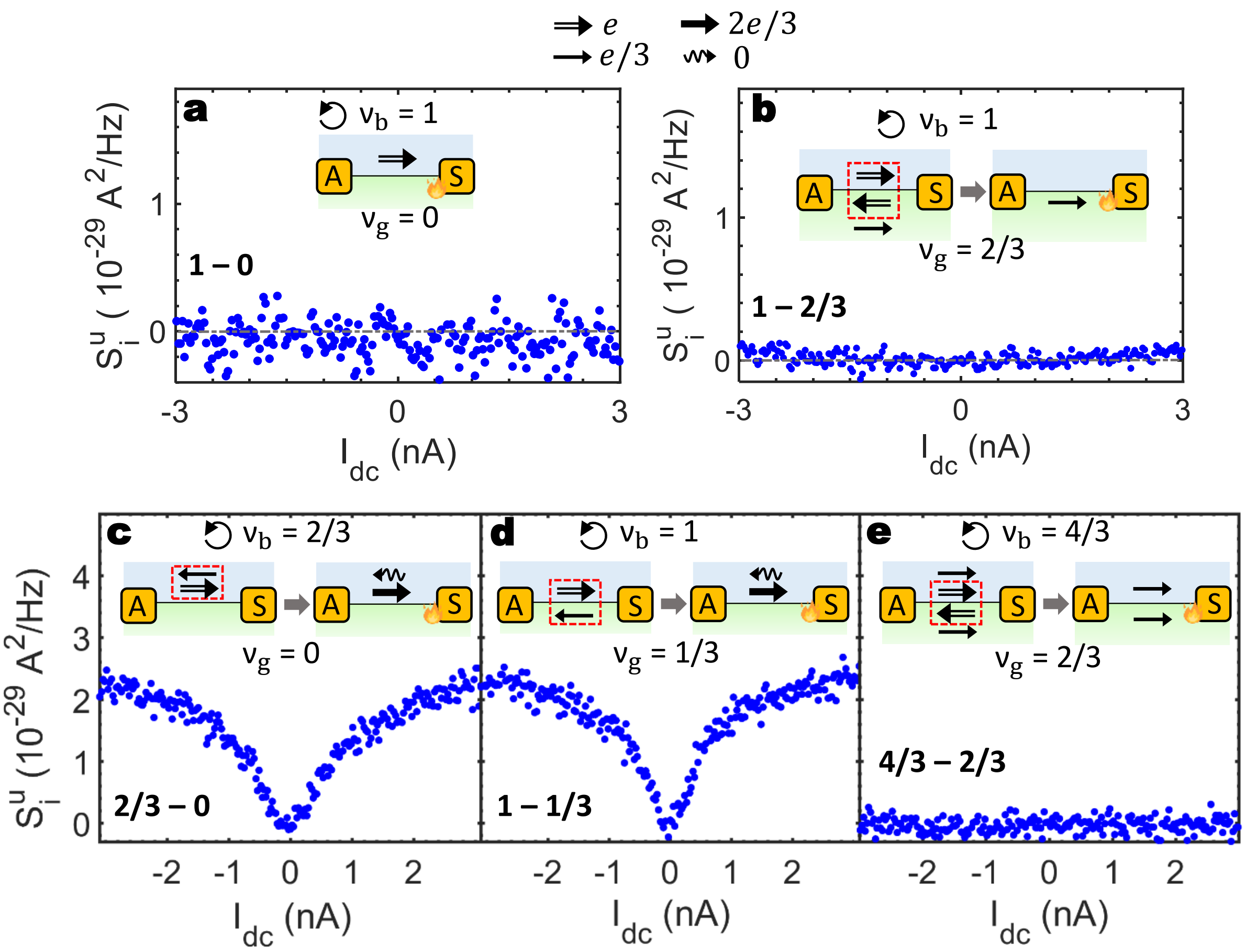

**ED Figure 8 | US noise measurements with interface edge modes. a,b,** Vanishing US noise ($S_i^u$) is measured for $1_b$-$0_g$=$1_{int}$ and $1_b$-$2/3_g$=$1/3_{int}$ modes at 70μm distance from the hot spot. **c-e**, $S_i^u$ of differently interfaced 2/3 edge modes: $2/3_b$-$0_g$, $1_b$-$1/3_g$ and $4/3_b$-$2/3_g$. Constituent modes and their equilibration are shown with symbolic arrows in the inset. A DC current generates hot-spot (shown by 'yellowish-fire') at the US side of the source contact, S. Full equilibration at $2/3_b$-$0_g$ and $1_b$-$1/3_g$, excites the US neutral which is measured at the amplifier contact, A (c,d). The $4/3_b$-$2/3_g$ configuration leads to two copropagating 1/3 modes along the interface, and thus it does not carry neutral mode. No US noise is measured in (e).

## VIII. No shot noise on t=1/2 for the configuration of $4/3_b$-$2/3_g$=$2/3_{int}$:

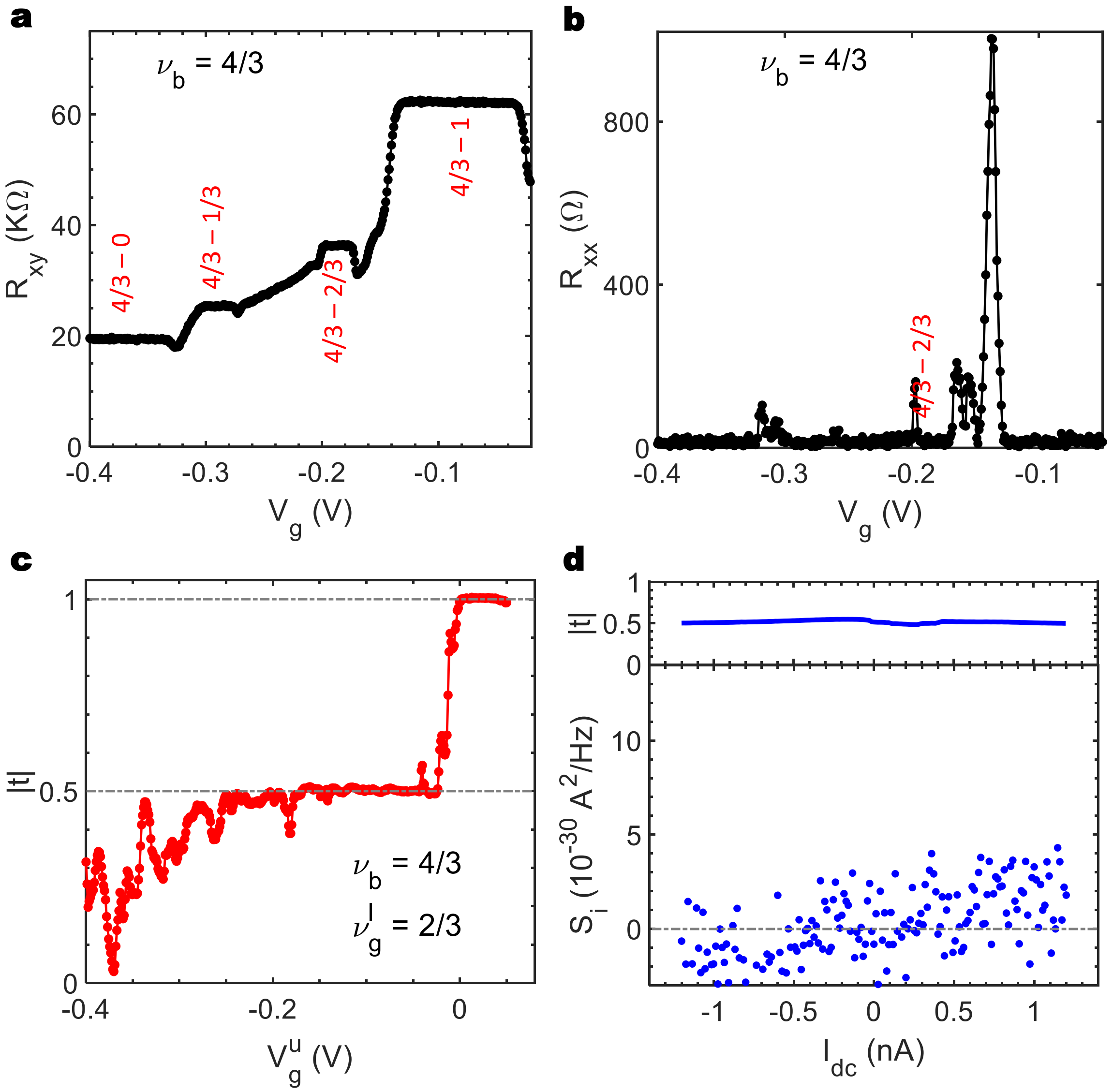

**ED Figure 9 | Results on $4/3_b$ bulk and $2/3_{int}$ edge. a,** Interface edge modes at $\nu_b = 4/3$ filling factor. **b,** Longitudinal interface resistance for $\nu_b = 4/3$. **c,** QPC transmission for $4/3_b$-$2/3_g$=$2/3_{int}$ mode showing two DS 1/3 modes and a conductance plateau t=1/2. **d,** Top: bias dependence of the transmission on the plateau of t = 1/2. Bottom: As expected, zero excess noise was measured in the DS amplifier.

The $4/3_b$-$2/3_g$=$2/3_{int}$ edge is free of neutral modes (see ED Fig. 1e) and thus free of shot noise (ED Fig. 8) on a conductance plateau. This novel engineered 2/3 edge can be thought of as an exact fractional analogue of conventional $\nu = 2$ edge mode, as each DS $\frac{e^2}{3h}$ edge mode belongs to each of the two Landau levels. This fractional analogue is therefore a smoking gun of revealing fractional interference and the exchange statistics of anyons[7,8].

# IX. Extended Theoretical Analysis

## i. The general paradigm and underlying assumptions.

Our derivation of a universal noise Fano factor rests upon several assumptions, which are believed to generally hold true in the experiment. Two disparate mechanisms may account for noise at the drain: (i) partial back-scattering (also known as beam splitting) or (ii) charge-to-neutral-to-charge conversion. Importantly, a transmission plateau, the entire noise results from the latter mechanism, which consists of two steps:

1. At the transmission and the reflection legs emanating from the QPC, if there is more than one charge DS mode, their occupancies will be different. The reason is that some DS moving charge modes emanate from a voltage biased source (cf. e.g., S1 of ED Fig. 11), while others originate from a grounded "source" (S2 of ED Fig. 11). These modes try to equilibrate to the same chemical potential while keeping the current conserved. As equilibration takes place (i.e., charge transfers from a highly occupied mode to a grounded one), neutral modes are generated as a by-product. (see ED Fig. 10).
2. These counter-propagating neutrals travel to the opposite side of the QPC geometry and then decays (cf. ED Fig. 11). While traveling US the identity of the neutral quasi-particles ("neutralons") is not protected[9], and may change from neutralon to anti-neutralon and vice-versa. Eventually, sufficiently far from the point of equilibration (where neutralons are generated), they become fully randomized, i.e., the probabilities for a neutral quasi-particle to be a neutralon or an anti-neutralon become equal. As such a quasi-particles decays, it gives rise to a quasi-particle/quasi-hole pair in the corresponding charge channels (and, with equal probability, to a quasi-hole/quasi-particle pair). A sequence of many such random events is manifest as a dc noise in a given charge channel, but vanishing dc current.

This paradigm depends crucially on the presence of the counter-propagating neutral modes. Here, we make a distinction between particle-like and hole-like fractional QH phases. For the former class, topology dictates only DS charge modes. By contrast, for hole-like phases, topology implies that US and DS modes co-exist. For particle-like fractions, US modes may appear only in the presence of edge reconstruction[3,7,10] (e.g. bulk filling fraction 1 with a reconstructed $1/3$ strip[10]). For hole-like fractions, US neutral modes could be the direct result of topological constraints that dictate the presence of counter-propagating edge modes: They may emerge as a consequence of the interplay of (renormalized) edge disorder and inter-mode interactions[5]. However, even this class of bulk phases exhibits edge reconstruction, as evidenced by a recent experiment[7]. The RG fixed points of the corresponding model[11] lead the emergence of charge modes moving DS and only US neutrals, whose origin is both the underlying topology and edge reconstruction. One such case is depicted in ED Fig. 10 with an example of inter-mode charge hopping in terms of the bare modes and the renormalized modes (the latter shows the creation/annihilation of neutralons). Our subsequent analysis relies on the following observations:

1. For fractional QH phases, US neutrals are always present, either due to topological constraints (for hole-like fractions) or as a result of edge reconstruction.
2. There are several important length scales in the problem: (i) The cutoff scale $\ell_{\mathrm{RG}}$ of our RG procedure. We assume here that on this scale we are sufficiently close to the RG fixed point; (ii) $\ell_{\mathrm{c,eq}}$, the charge equilibration length among the various charge modes; (iii) $\ell_{\mathrm{n,\bar{n}}}$, the length scale over which neutralons and anti-neutralons lose their identity and freely convert into each other; We assume that $\ell_{\mathrm{c,eq}}, \ell_{\mathrm{n,\bar{n}}} \ll L$, where $L$ represents the length of the edge segments emanating from the QPC.
3. Among the different possible charge equilibration processes only one is dominant; details of this single dominant process are not essential for our analysis.
4. While traveling from one side to the other side of the QPC, the neutralons completely lose their identity and become an equal mixture of neutralon and anti-neutralons.

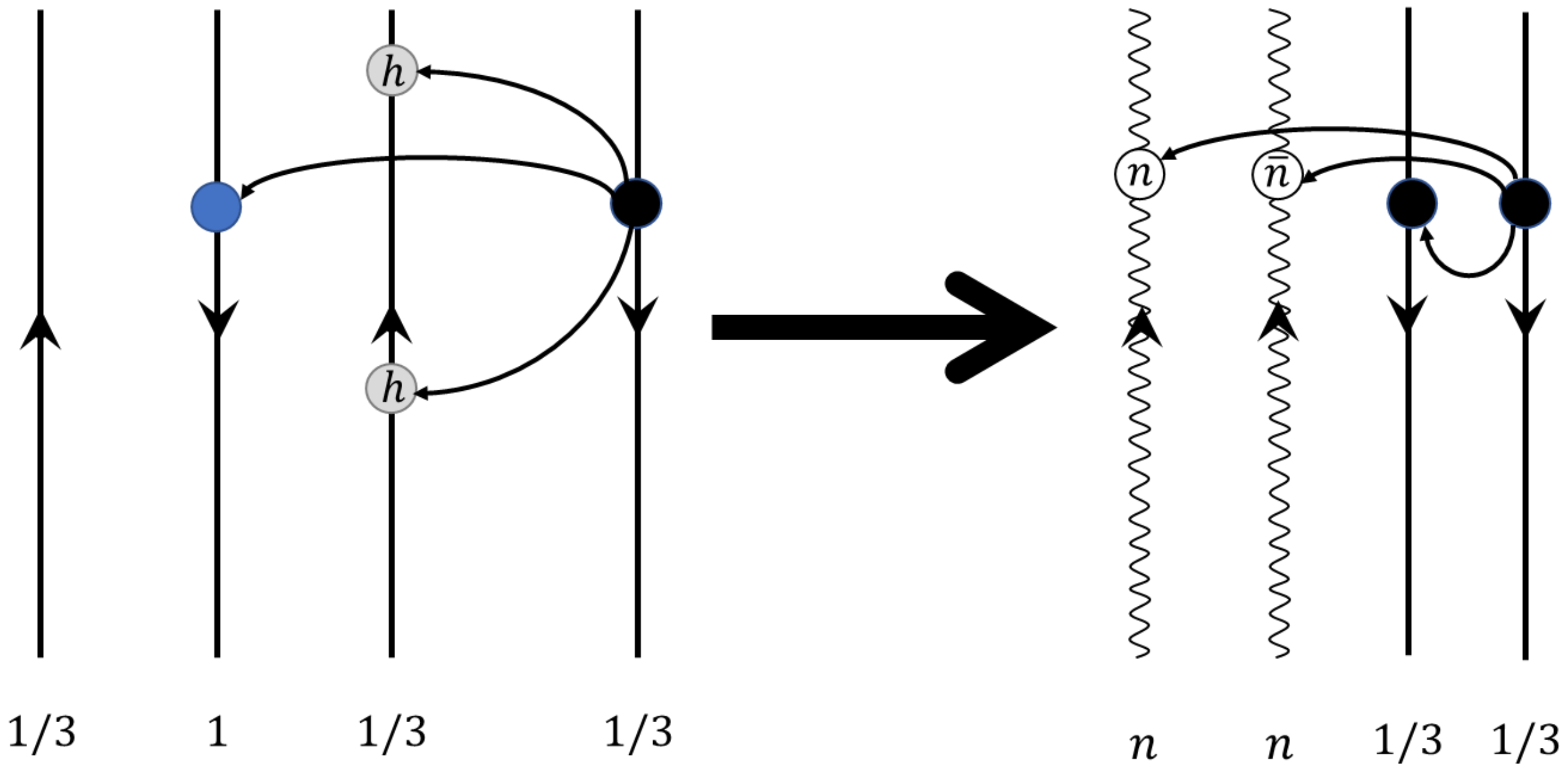

**ED Figure 10 | Charge equilibration on a reconstructed and renormalized edge. left,** we depict the bare reconstructed edge modes of a bulk phase $\nu = 2/3$. In the course of equilibration process, a single 1/3 quasi-particle (black circle) tunnels from the 1/3 DS outer mode (of higher chemical potential) to the inner DS 1 mode, creating a quasi-particle of charge 1 (blue circle) and two additional quasi-holes (grey circles) in the other $1/3$ edge mode[9]. **right,** the renormalized Wang-Meir-Gefen (WMG) edge[11]. The aforementioned process translates to a 1/3 quasi-particle from the outermost $1/3$ mode tunneling to the inner renormalized $1/3$ and creating, in addition, two US moving neutralons (white circles).

Below we address some representative case, showing the emergence of the universal law:

***Fano factor=Bulk filling fraction.***

### ii. Universal Fano factor: Particle-like bulk phases.

As we stated above, for integer or particle-like fractional bulk filling fractions there are no topological neutral modes. For such bulk phases neutral modes may emerge only due to edge reconstruction. It has been noted that even integer bulks may have fractional reconstructions[7,10]. Consider first the reconstruction of a $1/m$ strip near the edge of $2/5$ bulk. The interplay of inter-mode electrostatic interactions and disorder-induced inter-mode tunneling (between, e.g., two counter-propagating modes) may be accounted for within the framework of a renormalization group (RG) analysis, resulting in a fixed point underlain with three different DS modes of charges $e/15$, $e(m-3)/3m$, and $e/m$ (going from inner to outer-mode), along with an US neutral (see ED Fig. 11). While in this figure the neutral mode is presented as the inner-most mode, its precise location is of no consequence for our analysis.

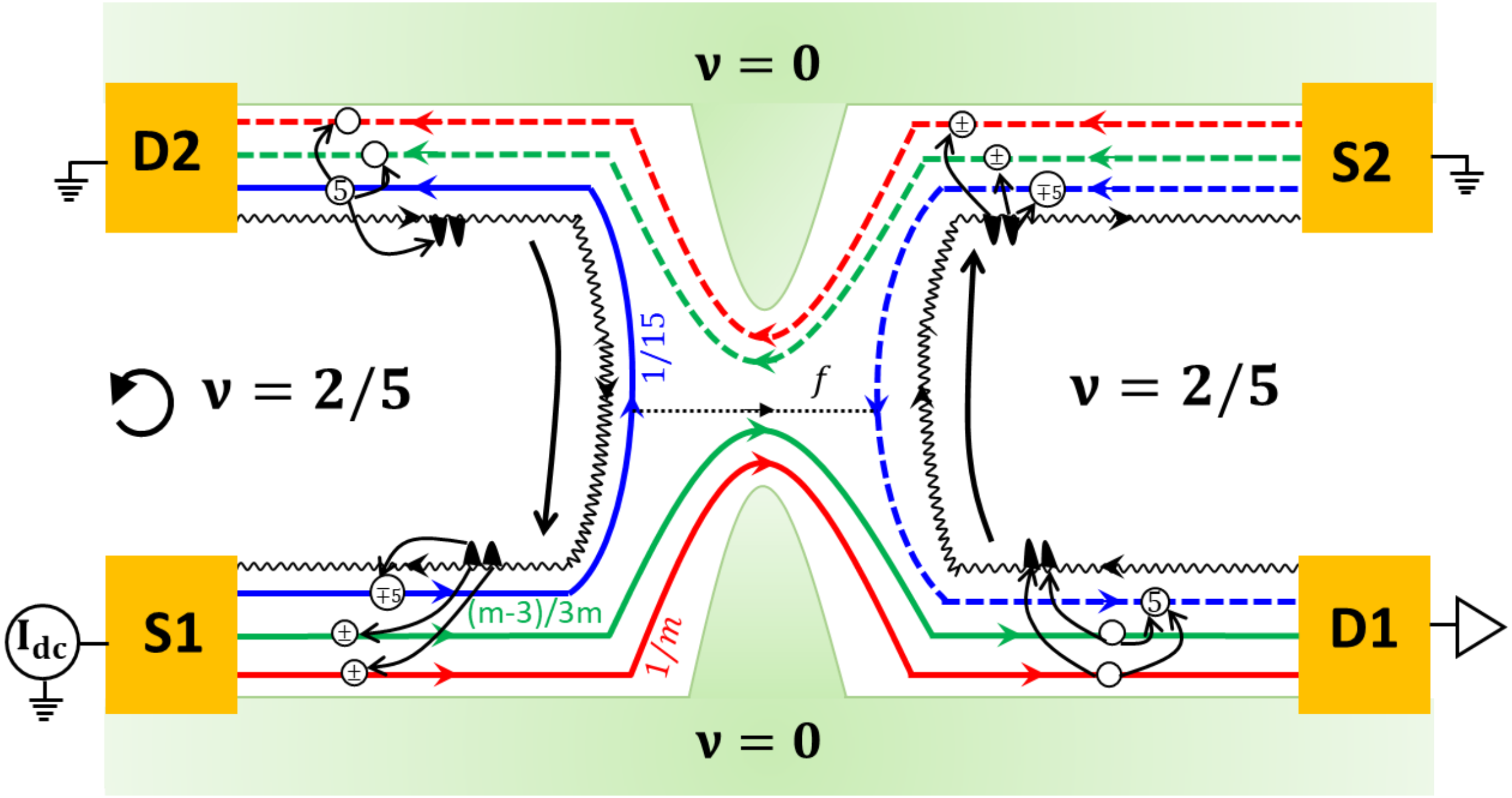

**ED Figure 11 | 2/5 edge with a QPC near the 1/3 plateau.** "Hot" charge modes, emanating from the voltage biased source S1, are denoted by solid lines; "cold" charge modes emanating from the grounded source S2 are represented by broken lines. While charge modes are represented by straight lines, neutrals are denoted by wavy lines. The charges of the modes, going from the inner to the outer-most, are $e/15$ (blue), $e(m-3)/3m$ (green), and $e/m$ (red). $f$ is the fraction of the e/15 charge mode which is transmitted to D1; $f=0$ at the plateau. After crossing the QPC on their way from S1 to D1, the $e(m-3)/3m$ and $e/m$ hot modes equilibrate with the $e/15$ cold mode. Charge transfer from the hot to the cold modes is accompanied by the creation of neutralons (the process is depicted near D1 with arrows). These flow US across the QPC and follow on, propagating towards S2. On their way, away from the QPC, the neutral quasi-particles (now, with equal probabilities neutralons and anti-neutralons) decay into charge excitations (see near S2 decay for the neutral decay process), thereby creating noise at D1 and D2 (even on the QPC conductance plateau, $f=0$). A similar thing happens near D2 where neutralons are created with the charge

transfer from the hot to the cold modes (depicted near D2). These neutralons travel US across the QPC towards S1 and decay to create charge excitations and hence noise at D1 and D2, resulting in a universal Fano factor for any $f$.

We next study the noise Fano factor near the $e^2/3h$ conduction plateau. We consider a time interval $\tau$ during which the source S1 injects N particles to each DS charge mode (we assume each mode carries charge at the same velocity). The total current injected from the source to the charge modes is then $I = 2eN/5\tau$. At the $e^2/3h$ conductance plateau ($f = 0$), the innermost $1/15$ mode is fully backscattered. As we move away from the plateau (cf. ED Fig. 11 with $f \neq 0$) the total charge reaching D1 over the time interval $\tau$ consists of the following contributions: $eN/3$ via the $(m-3)/3m$ and the $1/m$ modes which are fully transmitted, and $eNf/15$ via the partially transmitted $1/15$ mode. At the bottom right of Fig. S10, flowing towards D1, the $(m-3)/3m$, $1/m$ and $1/15$ modes are flowing in parallel. Accounting for the disorder-induced inter-mode tunneling, these modes equilibrate to assume the same chemical potential. This equilibration process comprises inter-mode charge transfer without affecting the total current arriving in D1. Following the equilibration, the number of quasi-particles carried by each mode over the time interval $\tau$ is $N_1$, with $N_1 = (5/2) \times (N/3 + Nf/15) = (5 + f)N/6$. Similarly, near the drain D2 we have a hot $1/15$ mode and cold $(m-3)/3m$, $1/m$ modes running in parallel. Conserving the total current, the number of quasi-particles carried by each mode, $N_2$ , is $N_2 = (5/2) \times (1 - f)/15 = (1 - f)N/6$. These two equilibration processes involve the creation of neutralons near D1 and D2, respectively. Those excited neutralons then travel to S2 and S1, respectively. As they run US, they lose their identity, becoming an equal mixture of neutralons and anti-neutralon. As these neutralons arrive near S1 or S2, they decay into fully randomized particle-hole pairs – as stated above, the corresponding edge segments (lower left and upper right in ED Fig. 11) are sufficiently long to facilitate full neutralon decay. This neutralon/anti-neutralon decay generates noise but *no dc current*. Let us denote the randomized pulses in a given mode near S2 and S1 by $a_i^{(\alpha)}$ and $b_j^{(\beta)}$, respectively, which assume the values $\pm 1$ for a quasi-particle/quasi-hole with equal probability. Here $\alpha, \beta$ denote the mode number running from inner to outer; $i, j$ run through the chronological sequence of pulses created. As we follow the decay of neutral quasi-particles we note that the process of a neutralon decay is precisely the inverse of the process of its creation. In this inverse process, a pair of neutrals decay to yield one quasi-hole in the $(m-3)/3m$ mode, one quasi-hole in the $1/m$ mode, and 5 quasi-particles in the $1/15$ mode. The inverse sign (quasi-particles↔quasi-holes) applies when anti-neutralons decay. Consequently, the number of particle-hole pairs is determined by the number of quasi-particles tunneling from one mode to the other during the process of charge equilibration. Thus, near S2 the number of excitations created in the three modes inner-to-outer is $N_1 - Nf, N - N_1, N - N_1$ respectively. Similarly, near S1 the number excitations will be $N(1-f) - N_2, N_2, N_2$ respectively. Due to the tunneling at the QPC, only a fraction $1 - f$ of the quasi-particles generated in the $1/15$ mode near S2 reaches D1. Similarly, given this tunneling bridge an extra fraction $f$ of noise in mode $1/15$ created near S1 reaches the drain D1. We can thus express the charge arriving at drain D1 as

$$Q_{\mathrm{D1}} = \frac{(5+f)Ne}{15} + \frac{e}{15}\sum_{i=1}^{(N_1-Nf)(1-f)} a_i^{(1)} + \frac{e}{15}\sum_{i=1}^{(N(1-f)-N_2)f} c_i^{(1)} + \sum_{i=1}^{N_2}\left[\frac{(m-3)e}{3m} b_i^{(2)} + \frac{e}{m} b_i^{(3)}\right]. \quad (1)$$

Here $a$, $b$ randomly assume the values $\pm 1$ ($b_i^{(2)}$ and $b_i^{(3)}$ are positively correlated as they are created together) with equal probability, $\langle Q_{\mathrm{D1}}\rangle = Ne/3$. $c_i$ are random variables (assuming the values $\pm 1$) that represent stochastic tunneling events due to tunneling across the QPC (which account for an additional noise term in Eq. 1). To compute the noise, we need to calculate the auto-correlation

$$\langle (Q_{\mathrm{D1}} - \langle Q_{\mathrm{D1}}\rangle)^2\rangle = \frac{f(1-f)Ne^2}{15^2} + \frac{(1-f)Ne^2}{45}. \quad (2)$$

We subsequently find the current-current auto-correlation at zero frequency to be

$$\ll I_{\mathrm{D1}} I_{\mathrm{D1}} \gg_{\omega=0} = \frac{2f(1-f)Ne^2}{15^2\,\tau} + \frac{2(1-f)Ne^2}{45\,\tau}. \quad (3)$$

Here $\frac{2f(1-f)Ne^2}{15^2\,\tau}$ is the noise due to the orthodox beam partitioning (tunneling across the QPC). Now using $I\tau = 2eN/5$ and $t = (5+f)/6$ ($t$ is the transmission parameter of the QPC) we calculate the effective Fano factor

$$F = \frac{\ll I_{\mathrm{D1}} I_{\mathrm{D1}} \gg_{\omega=0}}{2Ie \times t(1-t)} = \frac{2}{5}. \quad (4)$$

Remarkably, the Fano factor is equal to the bulk filling fraction. Importantly, the filling of the reconstructed edge strip ($1/m$) does not affect the value of the Fano factor, implying that even a small reconstruction (i.e., $m \gg 1$) can lead to a universal Fano factor. Moreover, we find that the Fano factor is independent of $f$, hence retains its universal value for $f = 0$, i.e. on the plateau. A similar calculation can be performed for all particle-like phases with edge reconstruction.

### iii. Universal Fano factor: Hole-like bulk phases

We can similarly calculate the noise Fano factor for any Abelian hole-like state (e.g. $2/3, 3/5, 4/7$) at any plateau. The general scheme is similar; we will thus present only one more example, that of bulk filling $2/3$ at the $1/3$ intermediate plateau, which follows the WMG[11] edge reconstruction picture (cf. ED Fig. 12). At the $e^2/3h$ conduction plateau we find the number of quasi-particles involved in equilibration near D1 ($N_1$) and D2 ($N_2$) to be $N_1 = N_2 = N/2$. The number of neutralon excitations is directly related to the number of tunneling quasi-particles involved in the inter-mode equilibration. We thus find that the total charge reaching the drain D1 is

$$Q_{\mathrm{D1}} = \frac{Ne}{3} + \frac{e}{3}\sum_{i=1}^{N_1} a_i^{(1)} + \frac{e}{3}\sum_{i=1}^{N_2} b_i^{(2)}, \tag{5}$$

where $a_i^{(1)}$, $b_i^{(2)}$ represent are random $\pm 1$ variables; here 1, 2 denote, respectively, the inner and outer mode, $i$ counts the chronological sequence of events, and $a$, $b$ represent different neutral decays near S2 and S1, respectively. This facilitates calculation of the current-current auto-correlation. Evaluating $\langle (Q_{\mathrm{D1}} - \langle Q_{\mathrm{D1}} \rangle)^2 \rangle$ we obtain $\ll I_{\mathrm{D1}} I_{\mathrm{D1}} \gg_{\omega=0} = \frac{2Ne^2}{9\tau}$. Now employing $I = 2Ne/3\tau$ and $t = 1/2$ we find

$$F = \frac{\ll I_{\mathrm{D1}} I_{\mathrm{D1}} \gg_{\omega=0}}{2Ie \times t(1-t)} = \frac{2}{3}. \tag{6}$$

A similar calculation for off-the-plateau correlations, along the lines of the $2/5$ case, yield the same Fano factor $F = 2/3$.

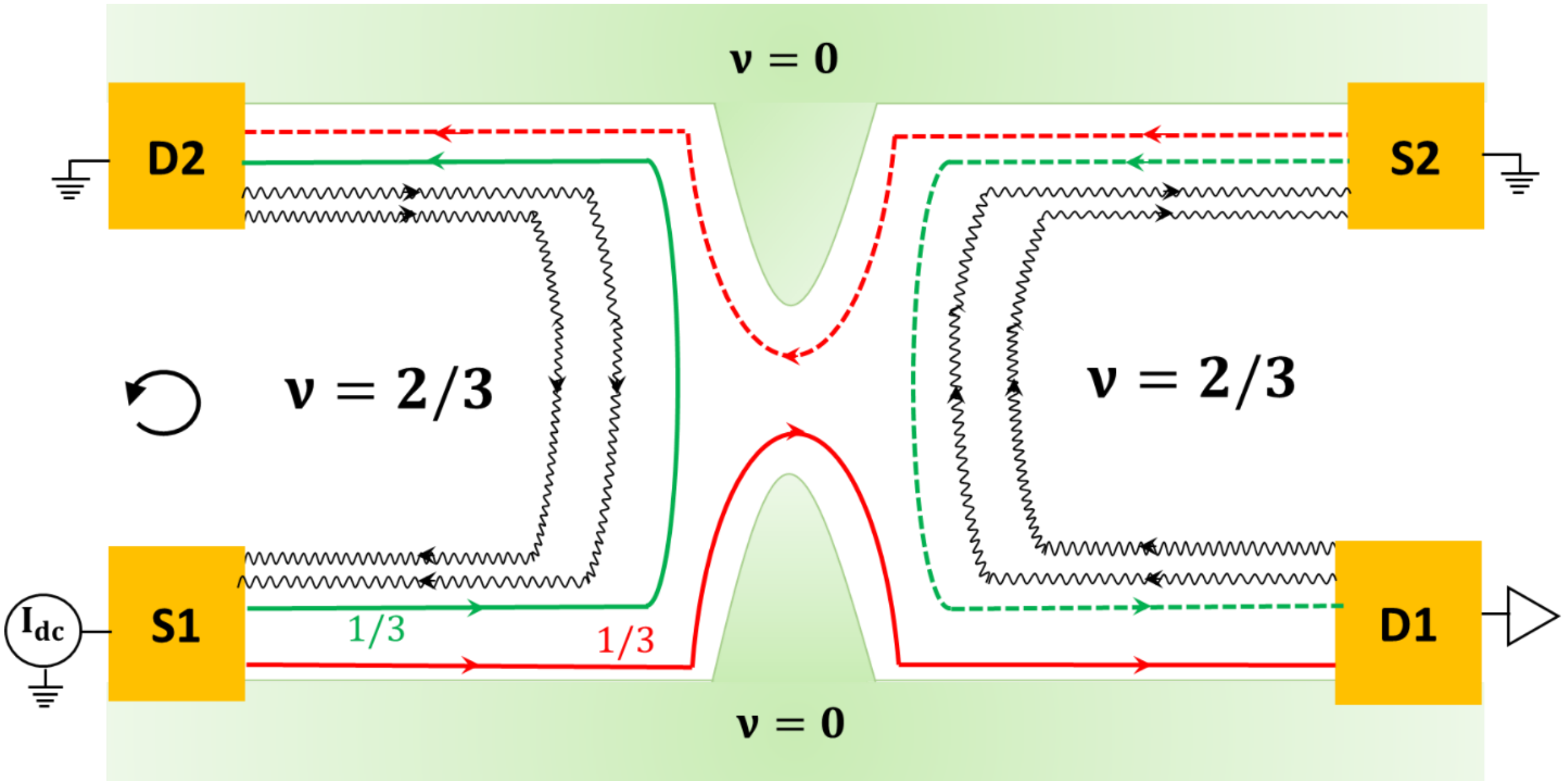

**ED Figure 12 |** Renormalized edge structure of the reconstructed edge of the 2/3 bulk, according to the WMG theory[11]. Pinching off of the QPC corresponds to the 1/3 conductance plateau. The edge modes comprise two US neutral modes (wavy black lines) and two DS 1/3 charge modes.

### iv. Universal Fano factor: General bulk filling

These calculations can be written for a general bulk filling fraction $\nu = \nu_1 + \nu_2 + \nu_3$, where $e\nu_1$ charge is transmitted to D1, $e\nu_3$ charge is fully reflected to D2, and a 1-$f$ fraction of $e\nu_2$ is transmitted to D1 (see ED Fig. 13). We can write the total charge reaching D1 as

$$Q_{\mathrm{D1}} = e(\nu_1 + (1-f)\nu_2) + e\nu_1 \sum_{i=1}^{(N-N_1)} a_i^{(1)} + e\nu_2 \sum_{i=1}^{(N_1-fN)f} c_i^{(2)} + e\nu_2 \sum_{i=1}^{((1-f)N-N_2)(1-f)} b_i^{(2)} + e\nu_3 \sum_{i=1}^{N_2} b_i^{(3)}. \quad (7)$$

This allows us to calculate the auto-correlation shot noise

$$\ll I_{D1} I_{D1} \gg_{\omega=0} = \frac{2Ne^2}{\tau}\left[\nu_2^2 f(1-f) + f\nu_1\nu_2 + \nu_1\nu_3 + \nu_2\nu_3 - f\nu_2\nu_3\right], \quad (8)$$

and the Fano factor is then

$$F = \frac{\ll I_{D1} I_{D1} \gg_{\omega=0}}{2Ie \times t(1-t)} = \nu_1 + \nu_2 + \nu_3 = \nu, \quad (9)$$

i.e., is given by the bulk filling fraction.

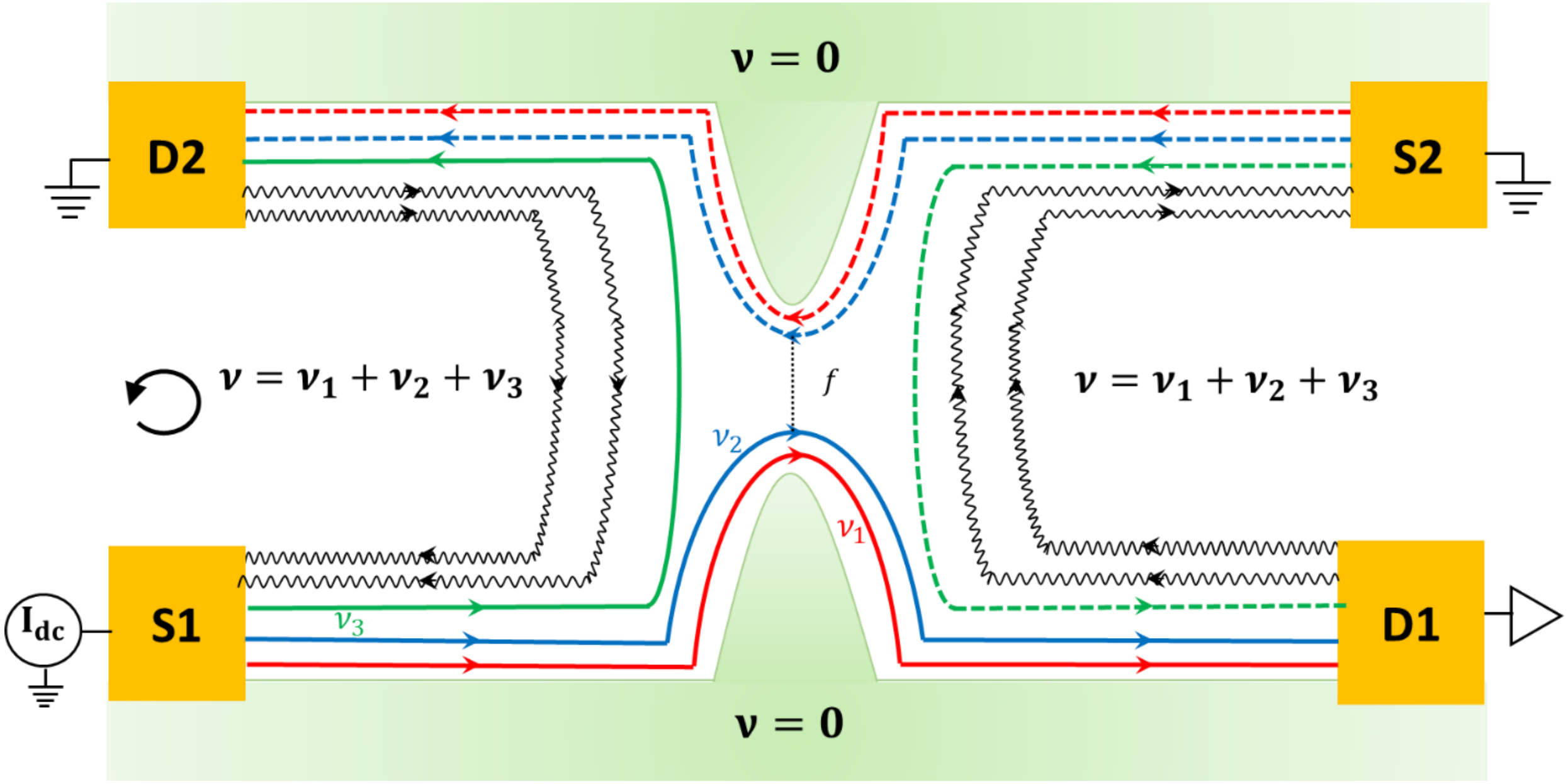

**ED Figure 13 | General bulk filling.** The QPC picture for the bulk filling $\nu = \nu_1 + \nu_2 + \nu_3$ at $[\nu_1 + (1-f)\nu_2]/\nu$ total transmission. The $\nu_3$ mode (green) is fully backscattered, $\nu_1$ mode (red) is fully transmitted and the $\nu_2$ mode (blue) is partially transmitted.

### v. Universal Fano factor: Engineered boundary.

A $1/3$ boundary fractional mode has been engineered by interfacing the $2/3$ phase with an embedding $1/3$ phase, leading to a $2/3$ edge (cf. ED Fig. 14). We note two important length scales: The charge equilibration length within the reconstructed 2/3 edge, $l_{\text{eq}}$, and the scale over which the two counter-propagating $1/3$ modes (from the 2/3 and the embedding 1/3 fractional QH phases) gap out, $l_{\text{gap}}$. We consider three different scenarios defined in relation to the edge segments leading to D1 and D2: (i) $l_{\text{gap}} \ll l_{\text{eq}}$: Charge equilibrates over a length scale $l_{\text{eq}}$ much longer than the one over which the outermost $1/3$ mode gaps out with the counter-propagating $1/3$ of the embedding 2DEG. (ii) These two length scales are comparable, $l_{\text{gap}} \sim l_{\text{eq}}$. (iii) $l_{\text{gap}} \gg l_{\text{eq}}$. When $l_{\text{gap}} \ll l_{\text{eq}}$, there is no possibility of equilibration near D1 or D2. Our charge-to-neutralon-to-charge mechanism is inactive, and only the orthodox protocol of shot noise generation through beam partitioning is effective. This yields a Fano factor equal to $1/3$ (associated with the charge of tunneling quasi-particles at the QPC). For $l_{\text{gap}} \sim l_{\text{eq}}$, both mechanisms for shot noise generation (beam partitioning and charge-to-neutral-to-charge conversion) are active, resulting in a non-universal Fano factor. In the regime $l_{\text{gap}} \gg l_{\text{eq}}$ equilibration takes place before gapping out of the counter-propagating 1/3 modes occurs; our charge-to-neutral-to-charge mechanism is then active, similar to the case of a boundary with a vacuum, cf. ED Fig. 13. The number of quasi-particle tunneling events in the course of equilibration near D1 and D2 is

$$N_1 = \frac{(1-f)N}{2}, N_2 = \frac{Nf}{2}, \tag{10}$$

respectively. In similitude to the analysis of the above cases, we compute the shot noise (due to both orthodox beam partitioning *and* charge-to-neutral-to-charge conversion):

$$\ll I_{D1} I_{D1} \gg_{\omega=0} = \ll I_{D1} I_{D1} \gg_{\text{ortho},\omega=0} + \frac{2e^2[((1-f)N - N_1)f + (Nf - N_2)(1-f)]}{9\tau}. \tag{11}$$

Employing $I = Ne/3\tau$, $\ll I_{D1} I_{D1} \gg_{\text{ortho},\omega=0} = \frac{2e^2(1-f)f}{9\tau}$ and $t = 1 - f$ we find,

$$F = \frac{\ll I_{D1} I_{D1} \gg_{\omega=0}}{2Ie \times t(1-t)} = \frac{2}{3}. \tag{12}$$

Hence the Fano factor is again universal if $l_{\text{gap}} \gg l_{\text{eq}}$.

A similar analysis has been performed for interfacing the $2/3$ phase with an embedding $4/15$ phase, leading to a $2/3 - 4/15 = 2/5$ edge, resulting in the same Fano factor.

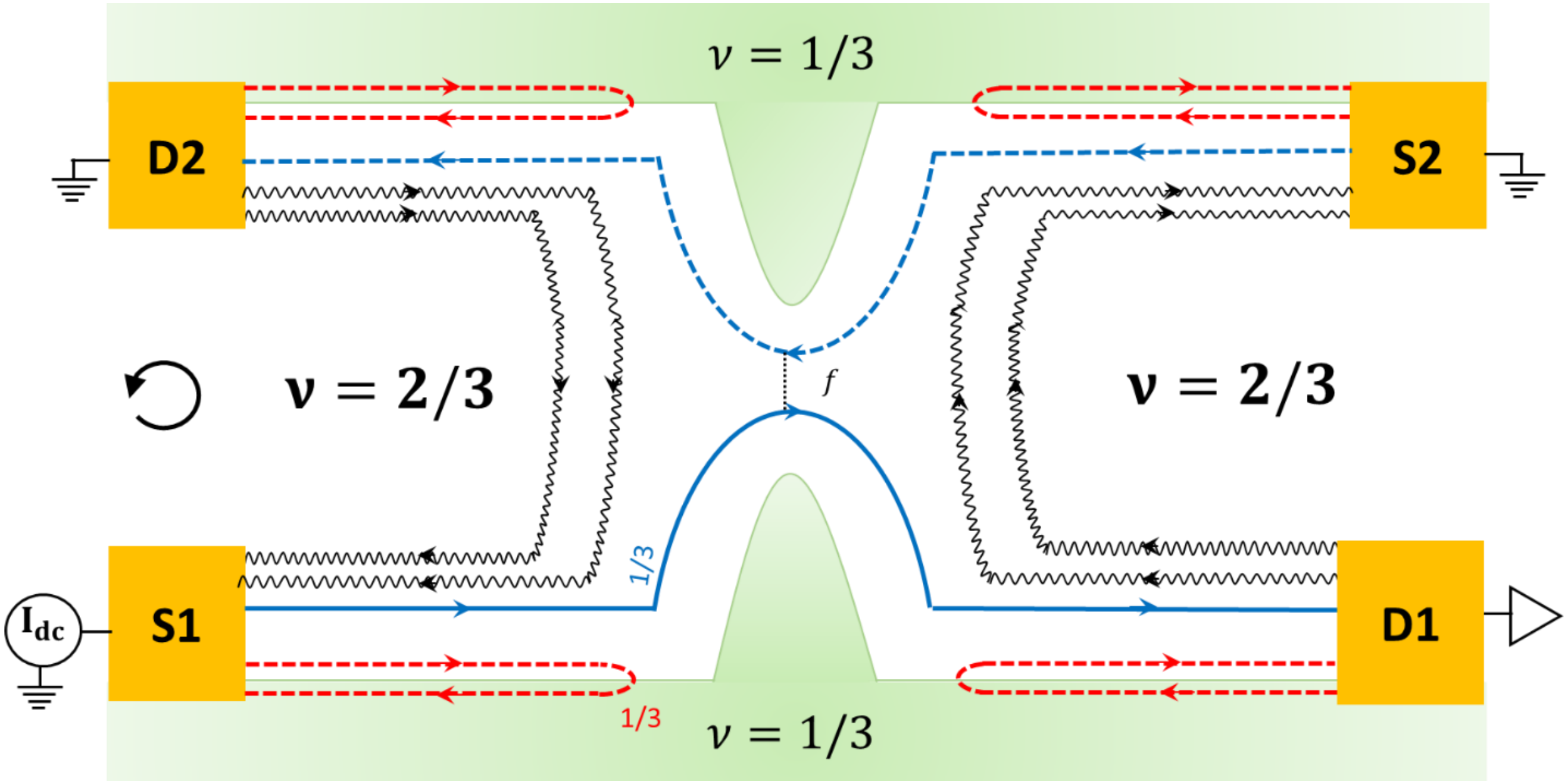

**ED Figure 14 | A 2/3 fractional QH phase embedded in a $\nu = 1/3$ fractional QH phase.** The dashed red line represents the gapped out outermost WMG mode with the counter-propagating

$1/3$ mode from the boundary of the embedding 2DEG. The ungapped inner $1/3$ mode is represented by a blue line. $f$ is the tunneling fraction of the inner $1/3$ mode.

### vi. Universal Fano factor: Non-equivalent engineered boundaries.

This interesting platform, shown in ED Fig. 15, comprises a bulk phase ($\nu = 1$) with two non-equivalent engineered boundaries: The top boundary is made up by interfacing the bulk phase with an embedding $\nu = 2/3$ phase, forming an artificial $1/3$ edge; likewise, the bottom boundary comprises interfacing with an embedding $\nu = 1/3$ phase, forming an artificial 2/3 edge. We will be focusing on the $1/2$ transmission plateau here. This edge structure of this plateau can be accounted if we postulate a $2/3$ reconstruction strip at the edge of a bulk $\nu = 1$ (cf. ED Fig. 15a). We also point out here that this has been often seen experimentally as a $2e^2/3h$ conductance plateau (see ED Fig. 15b). Following renormalization of this structure, the emerging $2/3$ and the neutral modes are then interfaced with the edge modes of the embedding 2DEG. At the 1/2 transmission plateau the inner $1/3$ mode is fully backscattered. Over a time interval $\tau$ the source S1 injects $N$ quasi-particles in each mode. However, as the DS $2/3$ of the bulk defined mode (cf. ED Fig. 15a) and counter-propagating $1/3$ mode of the embedding 2DEG have disorder induced tunneling, only $eN/3$ charge will reach the drain D1 giving us the $1/2$ transmission plateau. As we recognized before, here too there are two important length scales: $l_{\mathrm{RG}}$, over which the $2/3$ mode of the bulk and the counter-propagating $1/3$ mode of the embedding 2DEG (bottom boundary) get renormalized; and $l_{\mathrm{eq}}$, over which the renormalized co-propagating $1/3$ and $2/3$ modes of the bulk edge equilibrate. A universal Fano factor is achieved for $l_{\mathrm{eq}} \ll l_{\mathrm{RG}}$. In that case, the number of quasiparticles (associated with the bulk defined modes) near drain D1 that inter-mode tunnel owing to equilibration processes is $N_1 = N/2$. Similarly, near the D2 the equilibrated number of particles is $N_2 = N/2$. Thus, the number of quasi-particles/quasi-holes created in the bulk defined $1/3$ and the $2/3$ modes is $N_1$ and $N_2$, respectively. Owing to the fact at the bottom interface edge the bulk defined $2/3$ and the embedding 2DEG defined $1/3$ run anti-parallel to each other, only half the stochastic charge generated due to the annihilation of neutralons manifests as noise reaching the drain D1. Using the total transmission at the QPC $t = 1/2$ , and the total current injected from the source S1, $I = 2Ne/3$, we obtain that the total charge noise at the drain is

$$\langle\langle I_{D1} I_{D1} \rangle\rangle_{\omega=0} = \frac{2e^2}{3^2} N_1 + 2e^2 \times \frac{2^2}{3^2} \times \frac{N_2}{2} = \frac{Ne^2}{3}. \tag{13}$$

The Fano factor is then

$$F = \frac{\ll I_{D1} I_{D1} \gg_{\omega=0}}{2Ie \times t(1-t)} = \frac{Ne^2}{3} \times \frac{3}{4Ne^2} \times \frac{1}{\frac{1}{2}\left(1-\frac{1}{2}\right)} = 1, \tag{14}$$

in line with our general universal result, Fano factor=bulk filling fraction.

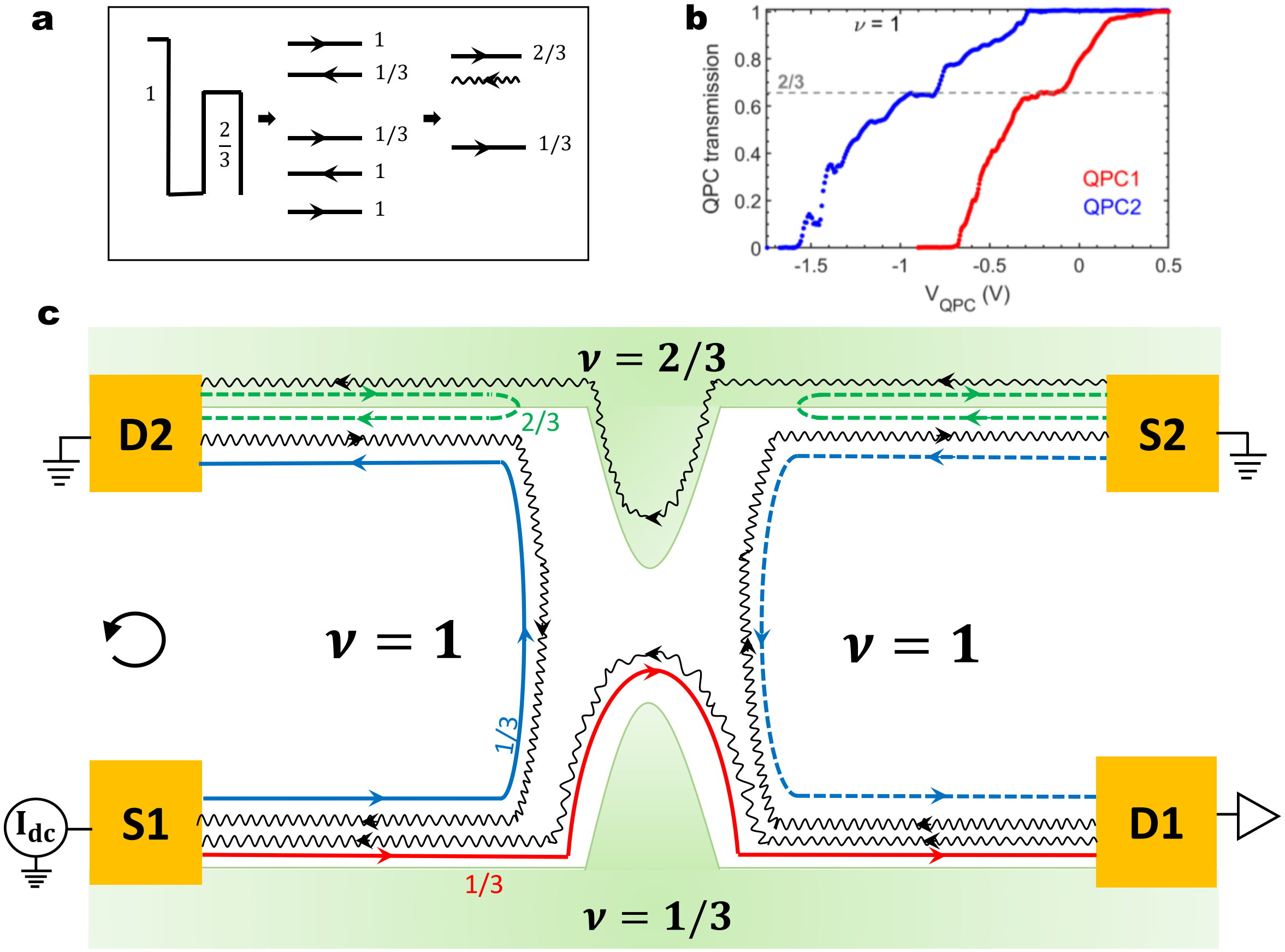

**ED Figure 15 | Edge structure of the bulk filling 1 with the bottom and the top gate at 1/3 and 2/3 filling, respectively. a,** Edge reconstruction at the ν=1 edge as a 2/3 strip. This reconstruction is needed to account for the half-transmission conductance plateau, seen in experiment (see ED Fig. 4). The RG fixed point gives rise to two DS charge modes, 2/3 and 1/3, and an US neutral. **b,** Observed 2/3 plateau at ν=1 for two different QPCs. **c,** This reconstructed

ν=1 edge mode is interfaced with the edge of a 1/3 fractional QH phase (bottom boundary) and 2/3 (top boundary), leading to the modes depicted in the figure. At the bottom edge we have shown the effective modes where the counter-propagating bulk defined DS $2/3$ and the embedding 2DEG defined US $1/3$ give rise to one DS $1/3$ mode (shown in red) and an US neutral. The blue mode is the other bulk defined 1/3 channel. At the top boundary the 2/3 green mode that is gapped out by the 2/3 mode of the embedding 2DEG (represented by the closed dashed green line). At the half-transmission conductance plateau the inner 1/3 mode is fully back scattered while the outer 1/3 mode is fully transmitted.

**References**

1 Bid, A., Ofek, N., Heiblum, M., Umansky, V. & Mahalu, D. Shot noise and charge at the 2/3 composite fractional quantum Hall state. *Physical Review Letters* **103**, 236802, doi:10.1103/PhysRevLett.103.236802 (2009).

2 Inoue, H. *et al.* Proliferation of neutral modes in fractional quantum Hall states. *Nature Communications* **5**, 4067, doi:10.1038/ncomms5067 (2014).

3 Khanna, U., Goldstein, M. & Gefen, Y. Emergence of Neutral Modes in Laughlin-like Fractional Quantum Hall Phases. *arXiv:2109.15293* (2021).

4 Bid, A. *et al.* Observation of neutral modes in the fractional quantum Hall regime. *Nature* **466**, 585-590, doi:10.1038/nature09277 (2010).

5 Kane, C. L., Fisher, M. P. & Polchinski, J. Randomness at the edge: Theory of quantum Hall transport at filling $\nu$ = 2/3. *Physical Review Letters* **72**, 4129-4132, doi:10.1103/PhysRevLett.72.4129 (1994).

6 Banerjee, M. *et al.* Observation of half-integer thermal Hall conductance. *Nature* **559**, 205-210, doi:10.1038/s41586-018-0184-1 (2018).

7 Bhattacharyya, R., Banerjee, M., Heiblum, M., Mahalu, D. & Umansky, V. Melting of interference in the fractional quantum Hall effect: Appearance of neutral modes. *Physical Review Letters* **122**, 246801, doi:10.1103/PhysRevLett.122.246801 (2019).

8 Goldstein, M. & Gefen, Y. Suppression of Interference in Quantum Hall Mach-Zehnder Geometry by Upstream Neutral Modes. *Physical Review Letters* **117**, 276804, doi:10.1103/PhysRevLett.117.276804 (2016).

9 Park, J., Rosenow, B. & Gefen, Y. Symmetry-related transport on a fractional quantum Hall edge. *Physical Review Research* **3**, 023083, doi:10.1103/PhysRevResearch.3.023083 (2021).

10 Khanna, U., Goldstein, M. & Gefen, Y. Fractional edge reconstruction in integer quantum Hall phases. *Physical Review B* **103**, L121302, doi:10.1103/PhysRevB.103.L121302 (2021).

11 Wang, J., Meir, Y. & Gefen, Y. Edge reconstruction in the $\nu$=2/3 fractional quantum Hall state. *Physical Review Letters* **111**, 246803, doi:10.1103/PhysRevLett.111.246803 (2013).